\documentclass{aa}
\usepackage{epsfig}
\usepackage{graphicx}
\usepackage{txfonts}
\begin{document}
   \title{Tracing the long bar with red-clump giants}

   \author{A. Cabrera-Lavers\inst{1,2},
           P.L. Hammersley\inst{1}, C. Gonz\'alez-Fern\'andez\inst{1}, M.
	   L\'opez-Corredoira\inst{1}, F. Garz\'on\inst{1,3}
	  \and
          T.J. Mahoney\inst{1}
          }

   \offprints{antonio.cabrera@gtc.iac.es}

 \institute{Instituto de Astrof\'{\i}sica de Canarias, E-38205 La Laguna, Tenerife, Spain\\
            \and
	    GTC Project Office, E-38205 La Laguna, Tenerife, Spain\\
                       \and
	 Departamento de Astrof\'{\i}sica, Universidad de La Laguna, E-38205 La
	 Laguna, Tenerife, Spain\\
                  }

   \date{Received XX; accepted XX}

 
  \abstract
   {Over the last decade a series of results have lent
   support to the hypothesis of  the existence of a  long thin bar in the Milky Way with a
   half-length of 4.5 kpc and a position angle of around 45$^\circ$. This is
   apparently a  very
   different structure from  
   the triaxial bulge of the Galaxy.}
   {In this paper, we analyse the stellar distribution in the inner 4 kpc
of the Galaxy to see if there is clear evidence
for two
   triaxial or barlike  structures, or whether there is only one. }
   {By using the red-clump population as a tracer of the
structure of the inner
   Galaxy we determine the apparent morphology of the  inner Galaxy. Star counts from  2MASS are used to  
provide
   additional support for this analysis.}
   {We show that there are two very different large-scale triaxial structures coexisting 
   in the inner
Galaxy: a long thin stellar bar constrained to the Galactic plane ($|b|<2^\circ$)
with a
   position angle of 43$\fdg$0$\pm$1$\fdg$8, and a distinct triaxial
bulge that extends to at least
   $|b|\le7.5^\circ$ with a position angle of 12$\fdg$6$\pm$3$\fdg$2. The scale height of the bar source distribution
 is around
   100 pc, whereas for the bulge the value of this parameter is five times larger.}
   {}


\authorrunning{A. Cabrera-Lavers et al.}
   \maketitle
%

\section{Introduction}

   There is now a substantial consensus for there being  a stellar bar in the
inner Galaxy.
   This  was first  suggested  by de
Vaucouleurs (1964) to explain the nonaxisymmetry in radio maps. The  first evidence for
a barlike distribution
in the stars  was derived from the asymmetries
in the infrared (IR) surface brightness maps  (e.g.\ Blitz \& Spergel 1991;
Dwek et al. 1995)
and in source counts  (Weinberg 1992; Hammersley et al.\ 1994;
Stanek et al.\ 1994),  which  both show systematically more stars at positive
galactic longitudes  $l<30$ close to the Galactic
plane (GP). The exact morphology  of the inner Galaxy, however, is still controversial. While some authors 
refer to the bar as a fatter structure, around 2.5
kpc in length with a position angle of 15--30 degrees with
respect to the Sun--Galactic Centre direction (Dwek et al. 1995;
Nikolaev \& Weinberg 1997; Stanek et al.
 1997; Binney et al. 1997;
Freudenreich 1998; L\'opez-Corredoira et al. 1999; 
Bissantz \& Gerhard 2002; Babusiaux \& Gilmore 2005), other researchers
suggest that there is a  long bar with a
half length of 4 kpc and a position
angle of around 45 degrees. It is noteworthy  that those authors supporting the $23^\circ$ bar all
examine the region  at $|l|<12^\circ$, whereas
those supporting the long bar with the larger angle are trying to explain counts for  
$10^\circ<l<30^\circ$ and $-10^\circ>l>-30^\circ$.

Hammersley et al.\ (2000) showed that there is a major over-density of
K2-3III (red-clump) stars
on the GP at \mbox{$l=27^\circ$} at a distance of about 6~kpc from the Sun. This over-density could also be
detected at smaller  galactic longitudes, but
 more reddened and at an increased distance from  the observer. These
stars  could not be seen  a few degrees above  
 or below the GP at the same galactic longitudes, nor at greater longitudes.
  This, when combined with the findings of Hammersley et al.\ (1994),  was interpreted as  
strong evidence for
 the existence of a long  in-plane bar with  a half length $\sim$4~kpc and
 position angle  of 43$^\circ$, clearly distinguishable from  the triaxial
bulge, but
running into it near $l$=12$^\circ$. This over-density of stars was also
analysed by
comparing near-infrared (NIR) counts with predictions of the
\emph{Besan\c{c}on Galactic model} (Robin et al.\ 2003)
with similar conclusions (Picaud et al.\ 2003).
 This result  has been further supported  by
observations in the mid-infrared with GLIMPSE data (Benjamin et al.\
2005) who did a similar analysis to that of 
Hammersley et al.\ (2000)  but at a wavelength range
in which the effects due to extinction are even lower than in the NIR, hence
reducing the uncertainty in the results.

There is also a noticeable difference in the luminosity function found in
the two regions. The majority of the work on the
short, fat bar (subtending an angle of $\sim$23$^\circ$ with the Sun--Galactic Centre line),  
has been restricted to within a few degrees of the plane in regions
that are routinely used by other authors to
study the bulge, and what is seen is an older population. The on-plane
region between $l=20^\circ$ and 27$^\circ$, however,  has been shown to contain a
large  number of extremely luminous young stars (Hammersley et al.\ 1994;
Gaz\'on et al.\ 1997;  L\'opez-Corredoira et al.\
1997), and  a characteristic feature of long bars is a major star formation region
where they interact with the disc.

These results suggest that the  structure in the inner Galaxy
($|l|<12^\circ$) could be interpreted as
a triaxial bulge\footnote{In other papers (L\'opez-Corredoira et al.\ 2000, 2005) we have
presented evidence that the Galactic bulge has a boxy morphology. The $(l,b)$ range examined in
the present paper, however, is too restricted to make a proper contribution to that debate; hence, for simplicity
we refer here to a triaxial geometry.} rather than a bar, and certainly not a long straight bar.  
 This  triaxial bulge has
 axial ratios of 1:0.5:0.4 (L\'opez-Corredoira
et al.\ 2005). Furthermore L\'opez-Corredoira et al.
(1999) showed that a triaxial bulge + disc model   cannot reproduce  
the observed counts in the Galactic plane
 for $10^\circ<l<30^\circ$. This is  also suggested by NIR photometry of 
red-clump stars (Nishiyama et al.\ 2005). It should be noted that the 
possibility of there being even
smaller non-axisymmetric structures (e.g.\ a secondary bar) in The Milky Way
cannot be discounted (Unavane \& Gilmore 1998; Alard 2001;
Nishiyama et al. 2005); thus, more observations are needed for a decisive conclusion to be reached.

Many galaxies have triaxial bulges contained within a primary stellar bar,
and even a third, smaller-scale bar inside the bulge (Fiedli et al.\  1996). More than
half of all disc galaxies have some kind of bar component (Buta, Crocker \&
Elmegreen 1996) and as many as a third have a strong primary bar (Freeman
1996). Recent NIR observations by  Beaton et al.\
(2005) and $N$-body simulations by Athanassoula \& Beaton (2006) provide strong
evidence in favour of the existence in our close neighbour M31 of a boxy bulge contained
within a primary bar of semilength of $\sim$4 kpc. Such a high degree of morphological similarity
between two neighbouring galaxies is striking (even though the angular separation between the major axes of
the primary bar and triaxial bulge is different for the two galaxies).

Bars are generally constrained to the
plane whereas bulges are shorter and more extended vertically.
Therefore,  using  data covering a wide range of longitudes and latitudes 
is the only way to distinguish between the  
two possible morphologies for the inner Milky Way: a short
inner bar or a triaxial bulge +  long bar. In this paper we have used NIR data from the  TCS-CAIN survey 
to analyse the spatial distribution of red-clump sources in
the inner Galaxy. This population of stars
 has been widely used as a standard candle in deriving distances to the
inner Galaxy (Hammersley et al 2000;
 Stanek et al. 1994; L\'opez-Corredoira et al. 2004;
 Babusiaux \& Gilmore 2005; Nishiyama et al.\ 2005, 2006b). By observing how
these stars are distributed along different lines of sight
towards the inner Galaxy   we are able to explore  the 3D morphology
of our Galaxy.


\section{TCS-CAIN catalogue}

TCS-CAIN catalogue is a NIR survey of the Milky Way that has 
recently been completed at the Instituto
de Astrof\'{\i}sica de Canarias. This survey consists of more
than 500 fields distributed along or near the
Galactic plane
($|b|\le10^\circ$) observed in $J$, $H$  and $K_{\rm s}$
filters with a photometric accuracy of $\sim$0.1 mag in the three bands
and a positional accuracy of $\sim$0.2$''$, based on the 2MASS catalogue as
astrometric reference. Each field
 covers approximately 0.07 deg$^2$ on the sky, centred on the stated  
galactic coordinates for that field. The
  nominal limiting magnitudes of the
 survey (determined using
 growth curves of the differential star counts) are 17, 16.5 and 15.2 in
$J$, $H$  and $K_{\rm s}$ respectively;
  hence, TCS-CAIN is at least one magnitude
 deeper than 2MASS, and even more in the inner Galaxy, where confusion is
important. A complete description
 of the catalogue and data
 reduction, and  
 details of its contents can be found in Cabrera-Lavers et al.\ (2006).


In this paper we have used all the available near plane fields in the inner Galaxy
($|l|<30^\circ$, $|b|$$\le$7.5$^\circ$), which gives a total number of 205 usable fields, 
covering an area of approximately  14.2 deg$^2$ 
of the sky (nearly 35\% of the whole catalogue area).

\section{Red-clump stars as a distance indicator}
Red-clump giants have long been proposed as standard candles. They have a
very narrow luminosity
function and constitute a compact
and well-defined clump in an HR diagram, particularly in the infrared. Furthermore, as they are relatively luminous, 
they can
be identified even to large distances from the Sun.
L\'opez-Corredoira et al.\ (2002) developed a method  
to obtain the star density and interstellar extinction along a line of
sight by extracting
the red-clump population from the NIR colour--magnitude diagrams (CMDs). This
method has proved very powerful for  analysing the
structure of the
 thin (L\'opez-Corredoira et al. 2002, 2004) and
thick (Cabrera-Lavers et al. 2005) discs of the Milky Way, as well as in
the study of the
 distribution of interstellar extinction (Drimmel et al.\ 2003; Picaud et
al.\ 2003; Duran \& van Kerkwijk 2006).


The  absolute magnitude ($M_K$) and intrinsic colour, $(J-K_{\rm s})_0$,
of the red-clump giants are well established
(Alves 2000; Grocholski \& Sarajedini 2002; Salaris \& Girardi 2002;
Pietrzy\'nski et al.\ 2003). Here, we consider an absolute magnitude
for the red
clump population of $-1.62\pm 0.03$ mag and and an intrinsic colour of
$(J-K_{\rm s})_0$ = 0.7 mag. These values are  consistent with the results derived by Alves
(2000) from the
{\itshape Hipparcos} red-clump and also with the results obtained by Grocholski \&
Sarajedini (2002)
for open clusters. The intrinsic colour of the red-clump stars
predicted by the
Padova isochrones in the 2MASS system (Bonatto et al.\ 2004) for a 10 Gyr
population of solar metallicity is $(J-K)_0$ = 0.68 mag. As 2MASS
photometry is
nearly equivalent to the TCS-CAIN photometric system (Cabrera-Lavers et al.\
2006) the value of
0.7 mag is sufficiently close to represent the red-clump population.

The absolute magnitude of the red-clump has a small dependence on metallicity
(Salaris \& Girardi 2002). There is no large metallicity gradient in
the  Galactic disc
 (Ibata \& Gilmore 1995a,b; Sarajedini et al.\ 1995); however,  there is  a
slight difference between the solar neighborhood metallicity and those
of the
bulge giants, which peaks at [Fe/H] = $-0.25$ dex (Ram\'irez et al.\ 2000;
Schultheis
et al.\ 2003; Molla et
al.\ 2000). According to Salaris \&
Girardi (2002), a small  correction of 0.03 mag should be applied to the
absolute
magnitude of the red-clump stars for these ranges of metallicity.
However,  this  is 
well within the uncertainty considered for the absolute magnitude used
here. The red-clump absolute magnitude in $J$ is more sensitive to
metallicity
and age than the $K$ filter, hence the intrinsic colour $(J-K)_0$ also
depends on both
the metallicity and age (Salaris \& Girardi 2002; Grocholski \& Sarajedini
2002; Pietrzy\'nski et al.\ 2003). The theoretical isochrones of Salaris
\& Girardi (2002)
and Bonatto et al.\ (2004) suggest  that  the de-reddened colour for the
range of
metallicities expected for a bulge population is $(J-K)_0=0.63\pm0.02$.
Hence the  metallicity dependences lead to a systematic
uncertainty of 0.03 mag in the absolute magnitude of the red-clump stars and
0.05 mag in the intrinsic colour, $(J-K)_0$.

A recent study using NIR spectra of the inner Galaxy giants  between  
$20^\circ<l<30^\circ$ has shown that these stars have a well defined 
metallicity (typically [Fe/H] between 0 and $-0.2$ dex) with a very small dispersion 
(Cabrera-Lavers et al., in preparation).
By using the population of infrared carbon stars from the 2MASS survey,
Cole \& Weinberg (2002) conclude that
the bar  formed more recently than 3 Gyr ago and must be younger than
6 Gyr. For this age and metallicity, the 
corrections given by   Salaris \&
Girardi (2002) are even lower than for the older bulge. Hence, we have
assumed that
 uncertainties in absolute magnitude and intrinsic colour for this
population of red-clump stars are of
  the same order as in the case of the bulge stars.

In this paper the TCS-CAIN survey data is analysed using the method
developed by Stanek et al.\ (1997) for
their data obtained by the Optical Gravitational
Lensing Experiment (OGLE; Udalski et al.\ 1993,1994). This  method
 has also recently  been used by Nishiyama et al.\ (2005, 2006b) and Babusiaux \&
Gilmore (2005) to analyse
the red-clump population of the inner Galaxy.

 To avoid extinction effects a reddening-independent
magnitude  ($K_e$) is used:

\begin{equation}
K_e=K_s-\frac{A_{K_{\rm s}}}{A_J-A_{K_{\rm s}}}(J-K_{\rm s}),
\label{ke}
\end{equation}
adopting $A_{K_{\rm s}}/E_{J-K_{\rm s}}=0.68$ (Rieke \& Lebofski 1985), a value
which agrees with that used in
 Babusiaux \& Gilmore (2005)
and also with that obtained by Indebetouw et al.\ (2005) by
combining data from the \emph{Spitzer Space Telescope} and 2MASS.

\begin{figure}[!h]
\centering
\includegraphics[width=8cm]{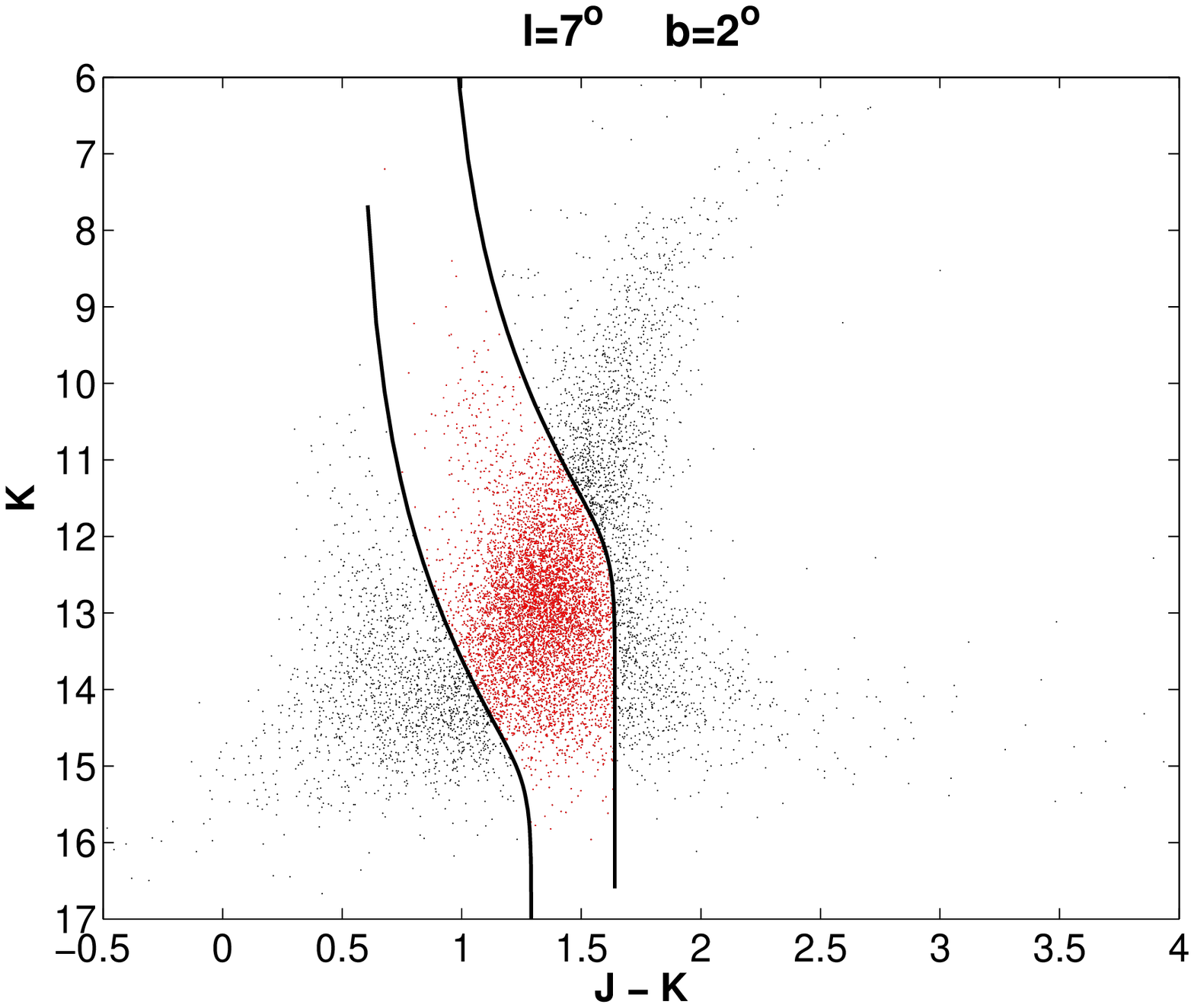}
\includegraphics[width=8cm]{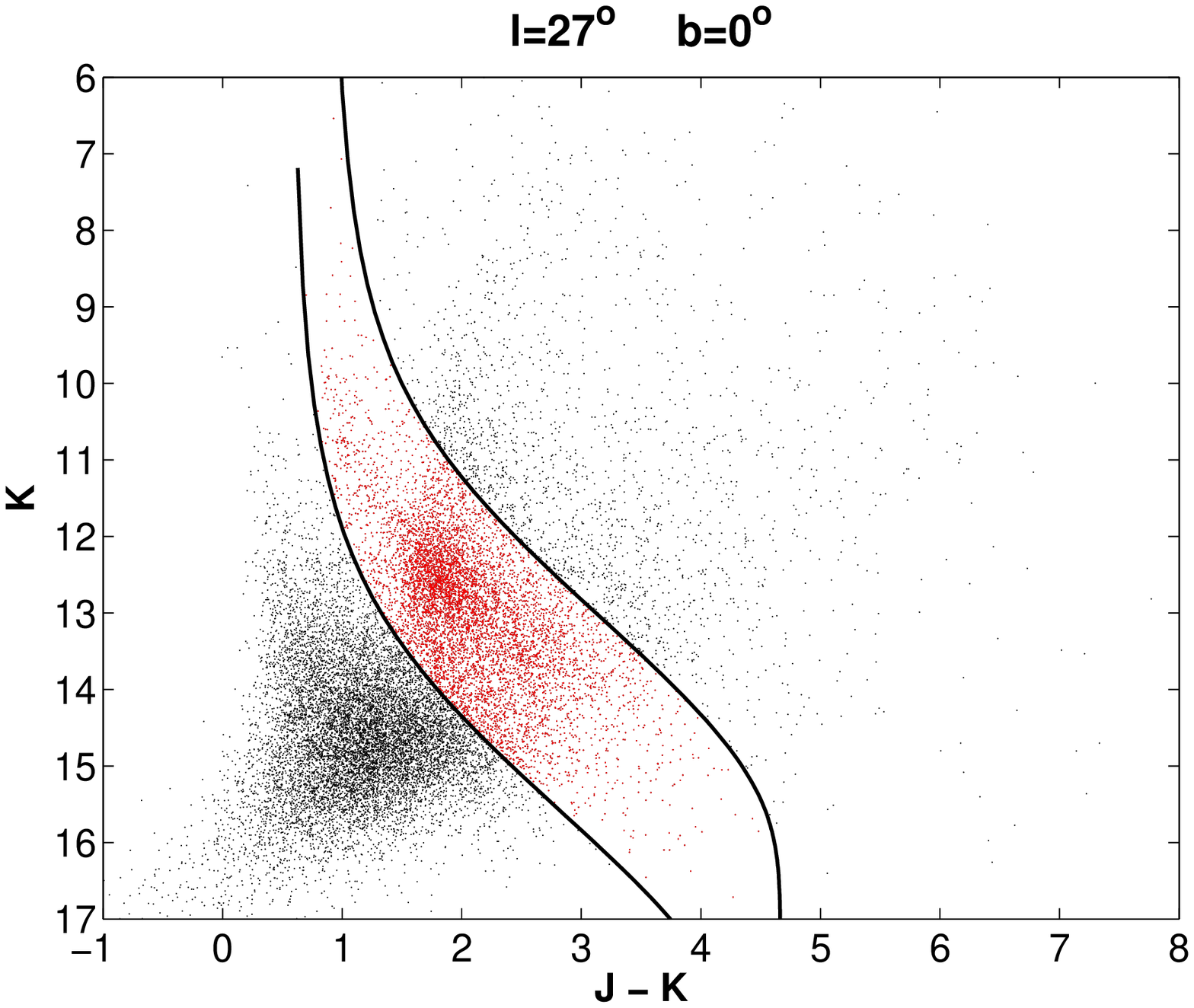}
\caption{Two examples of the use of the theoretical traces for the giant
population by means of the SKY model. Stars isolated between
both traces (corresponding to the K0 and the M0 populations) are
assumed to be red-clump stars and are extracted to analyse their spatial
distribution.}
\label{trazas}
\end{figure}

With the reddening-independent magnitudes, any compact structural feature in the CMD
would appear as a ``clump'' of stars, as they will be located at
approximately the
same distance from the Sun and hence they would have the same apparent
$K$-magnitude. As the red-clump stars are the dominant giant  population 
(Cohen et al.\ 2000; Hammersley et al.\ 2000), we can derive the
distance to
the feature by a non-linear least-squares fit  of the sum of a Gaussian
function
and a second-order polynomial (eq.\ \ref{nm}) to the histogram of red
giant stars (Stanek et al.\
1994; Stanek \& Garnavich 1998; Babusiaux \& Gilmore 2005):

\begin{equation}
N(m)=a+bm+cm^2+ \frac{N_{\rm RC}}{\sigma_{\rm RC}\sqrt{2\pi}}
\exp\left[-\frac{(m_{\rm RC}-m)^2}{2\sigma_{\rm RC}^2}\right].
\label{nm}
\end{equation}

The Gaussian term corresponds to a fit of the red-clump population of the
bulge/bar, while the second-order polynomial reflects the contribution
of the
background stars. The combined effect of distance plus extinction
produces an increase in
the  mixture between faint dwarfs and giants. In order to subtract the
contribution of foreground dwarf
stars we have used the SKY extinction  model (Wainscoat et al.\ 1992)
to derive the
theoretical traces in the diagrams for the giant population in each
field. With these theoretical
traces we can isolate the population of red-clump stars in each CMD;
thus, the
fit is made only   after removing the contribution of dwarf stars in the
reddening-corrected counts (Fig. \ref{trazas}).

To transform the derived extinction-corrected magnitudes into distances
from the
Sun we use:

\begin{equation}
\mu_K= K_{e} + \frac{A_{K_{\rm s}}}{A_J-A_{K_{\rm s}}}(J-K_s)_0 - M_K.
\end{equation}

To estimate an upper limit to the reddening-independent magnitude that
can be analysed reliably we
have followed the same procedure described in
Babussiaux \& Gilmore (2005). First, an estimate of the completeness
limits in $J$ and $K$ ($J_{\rm C}, K_{\rm C}$)
are derived by determining when the number
counts as function of magnitude stops increasing. The relative
completeness of $K_e$ is then derived by combining these estimates with the
mean colour of the giants in the field $(J-K)_{\rm M}$, and adding a 0.5 mag
safety margin to take into account the spread in the mean colour:

\begin{equation}
max(K_e)=min(J_{\rm C}-(J-K)_M, K_{\rm C})- \left(\frac{A_K}{E_{J-K}} (J-K)_{\rm M}
\right)+ 0.5.
\label{max}
\end{equation}

\begin{figure}[!h]
\centering
\includegraphics[width=8cm]{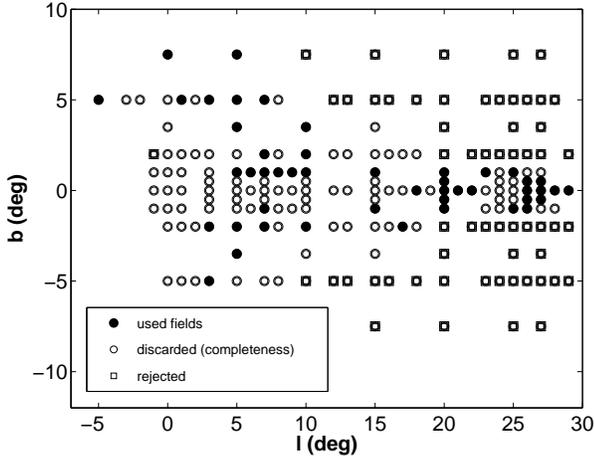}
\caption{Summary of the 205 TCS-CAIN fields analysed in this work.}
\label{campos}
\end{figure}

\begin{figure}[!h]
\centering
\includegraphics[width=8cm]{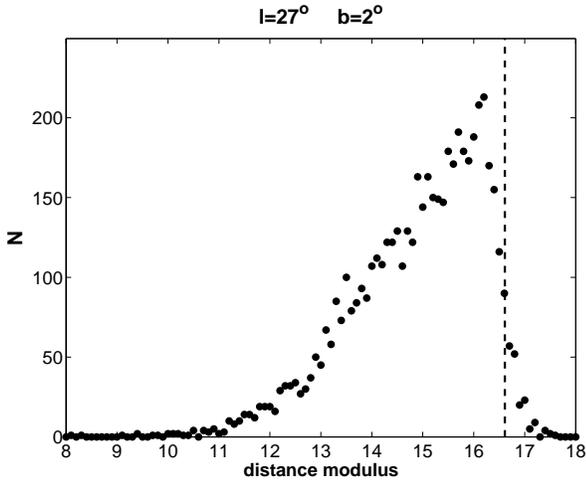}
\caption{Example of a field where no peak in the counts of  the red-clump distribution
is observed, as we are observing a field where only the disc component
is present, hence the star counts increase up to the completeness limit
of the field (shown as a
vertical dashed line).}
\label{figdisc}
\end{figure}

Figure \ref{campos} shows all of the analysed fields used in this work. There are
three
different kind of fields: 
\begin{itemize}
\item Those (labelled ``discarded'' fields) in which the completeness
made it impossible to fit  the red-clump
distribution accurately. 
\item  Those (labelled ``rejected'' fields) in which no
successful fit was possible, as no clear
structural component was
present. In these cases, the counts increase
continuously with the distance modulus until reaching
the completeness limit for the field. These in-plane fields are purely
disc fields and are located too far above or below the galactic plane for
 either the bulge or the bar components to be observed, so that no ``hump'' in
the red-clump counts is expected (Fig. \ref{figdisc}).
\item  Those  marked as ``used''  in Fig. \ref{campos}  where it was possible to  derive a 
distance to the red-clump population.  After discarding all those fields which were not 
useful, we obtained results in 49 of the total sample of 205
fields.
\end{itemize}

\subsection{Distances derived with $H/J-H$ data}
Some fields were observed only in the $J$ and $H$
bands, since far from the Galactic plane the counts in the $K$
and $H$ bands are nearly equivalent due to the low  extinction. For
this reason   some of the fields in this work use
$H$-band extinction-corrected magnitudes ($H_e$): 

\begin{equation}
H_e=H-\frac{A_{H}}{A_J-A_{H}}(J-H),
\label{he}
\end{equation}

\begin{equation}
\mu_H = H_{e} + \frac{A_{H}}{A_J-A_{H}}(J-H)_0 - M_H,
\end{equation}
using $(J-H)_0=0.5$, $M_H =-1.5$ (Wainscoat et al.\ 1992) and  
$A_H/E_{J-H}=1.41$, following Rieke \& Lebofsky (1985) and in agreement
with the  result of Nishiyama et al. (2006a). These values were used 
 in eight fields, those with  $|b|=3.5^\circ$ and 
$|b|=7.5^\circ$, and three additional fields at $b=5^\circ$ (those at $l<0^\circ$).

\begin{figure}[!h]
\centering
\includegraphics[width=8cm]{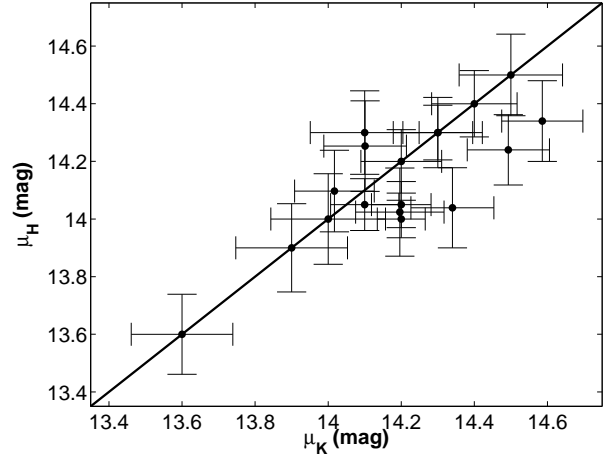}
\caption{Comparison between the distance modulus obtained with $H$-band
data with that obtained with $K$-band data for the same
fields. The solid line shows the 1:1 ratio. The two
quantities are well correlated and give similar distances.}
\label{figcompareH}
\end{figure}

\begin{figure}[!h]
\centering
\includegraphics[width=8cm]{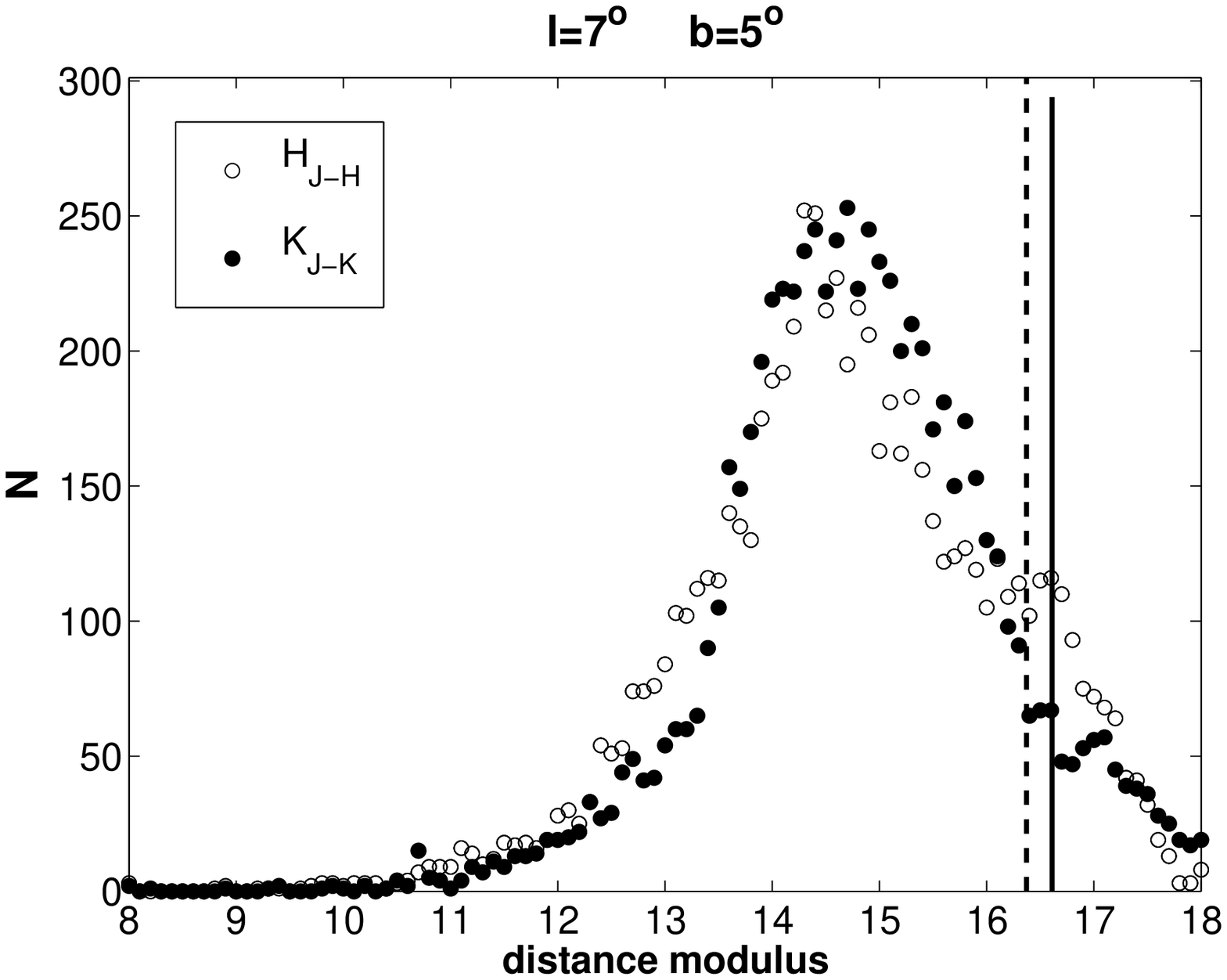}
\includegraphics[width=8cm]{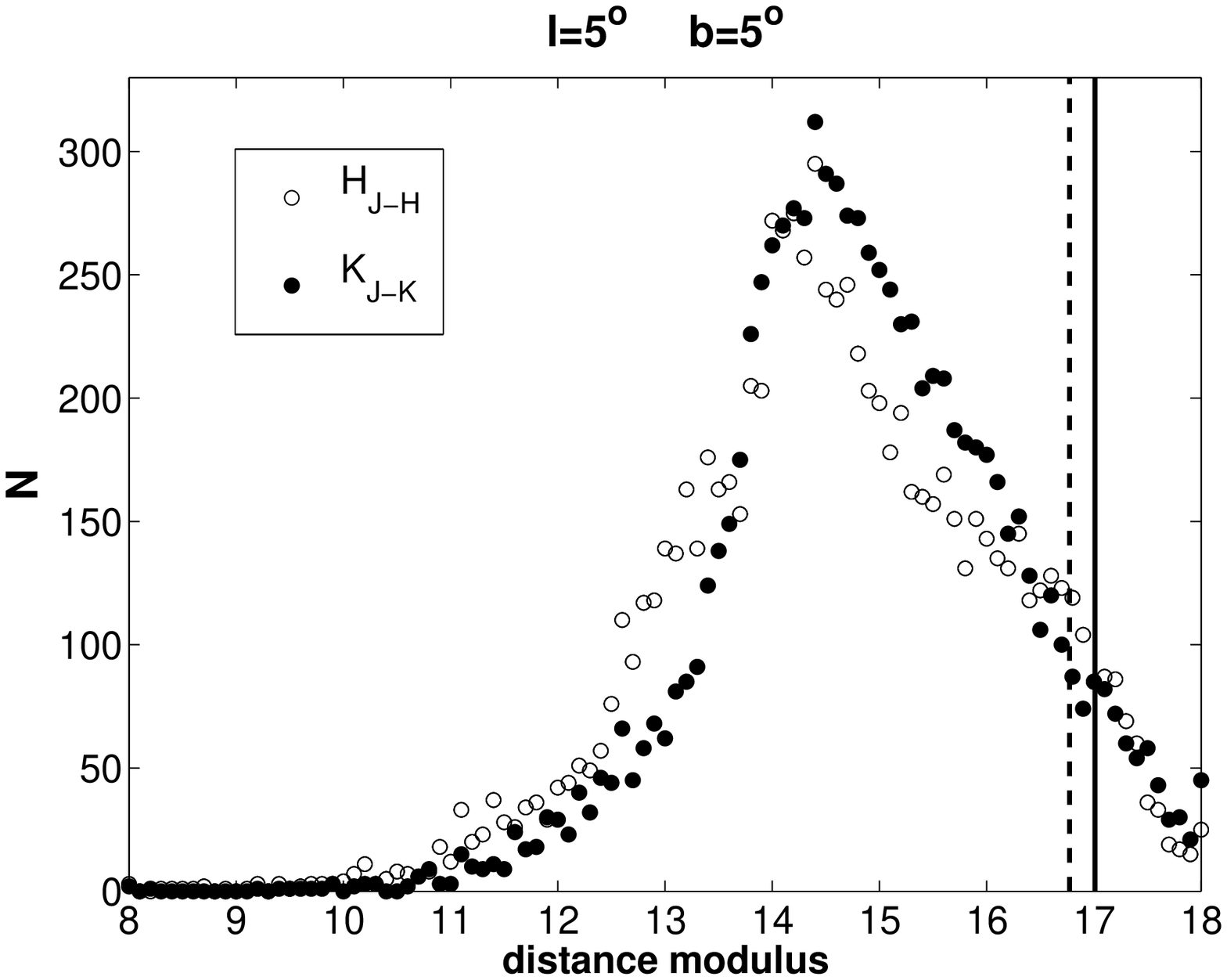}
\caption{Spatial distribution of red-clump stars obtained with either $H$-
or $K$-band corrected magnitudes for the
fields $l=7^\circ$, $b=5^\circ$ (above) and $l=5^\circ$, $b=5^\circ$
(below). The distributions are very similar, and the maxima are nearly
coincident. Vertical lines show the limiting magnitude derived using the $K$
(solid) and $H$ (dashed) bands,
respectively.}
\label{modulos}
\end{figure}

To test the reliability of deriving distances using either $K$- or $H$-band
corrected magnitudes, the results for
the fields with $2^\circ \le b \le 5^\circ$ were compared where 
 $JHK_{\rm s}$ photometry was available; also, there are no
effects of
completeness in either $H_e$ or $K_e$. In general, the fits using
the two different reddening-independent magnitudes are all consistent to
within 0.2 mag, with a mean difference of 0.05 mag ($\sigma$ = 0.11 mag)
for a total of 18 fields (see Figs. \ref{figcompareH} and
\ref{modulos}). Hence,  the 
derived distances using $(J-H,H)$ data can be compared with those obtained
using
$(J-K,K)$ data.

\section{Model fits}
\begin{figure}[!h]
\centering
\includegraphics[width=6.00cm]{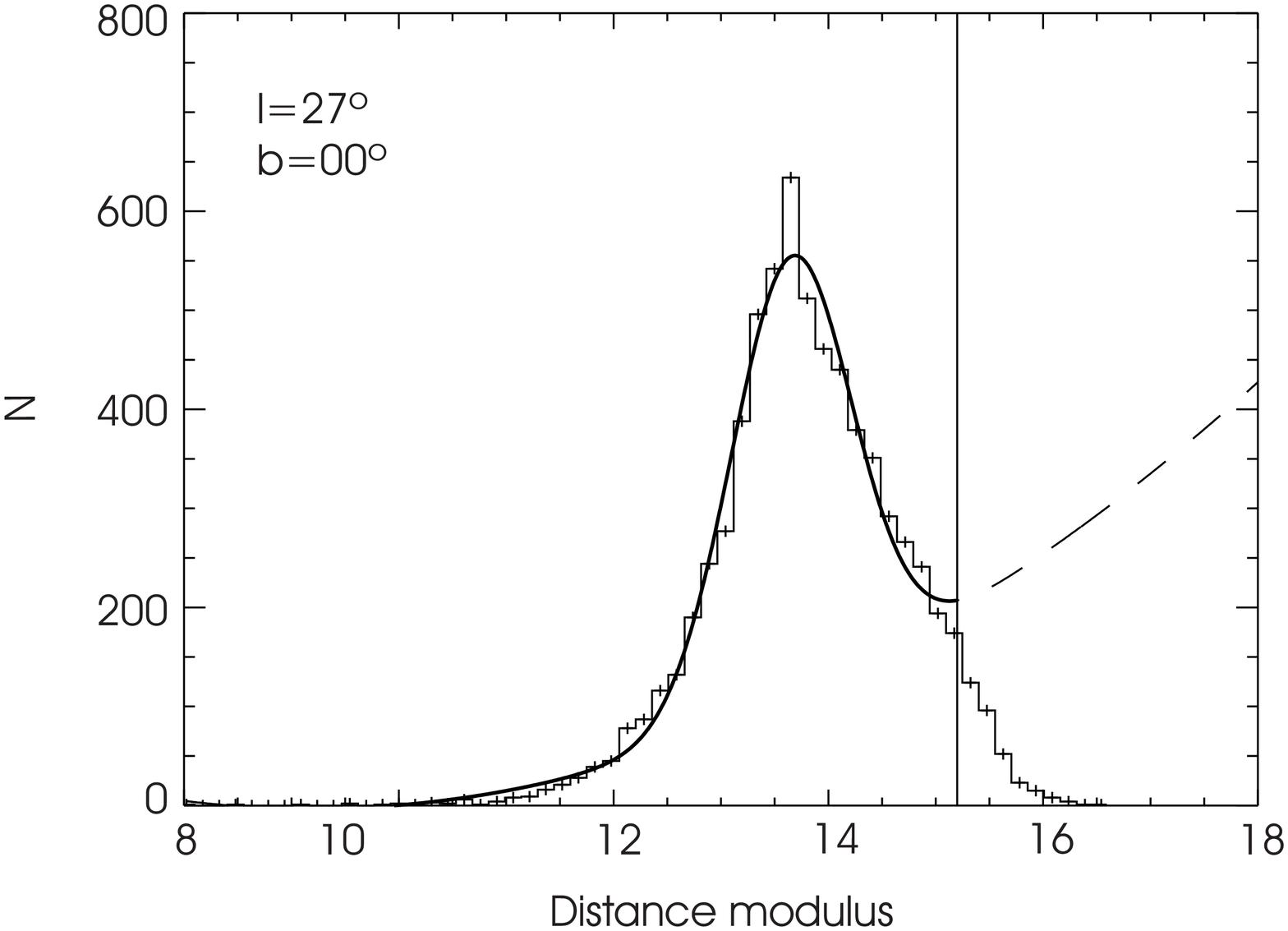}
\includegraphics[width=6.00cm]{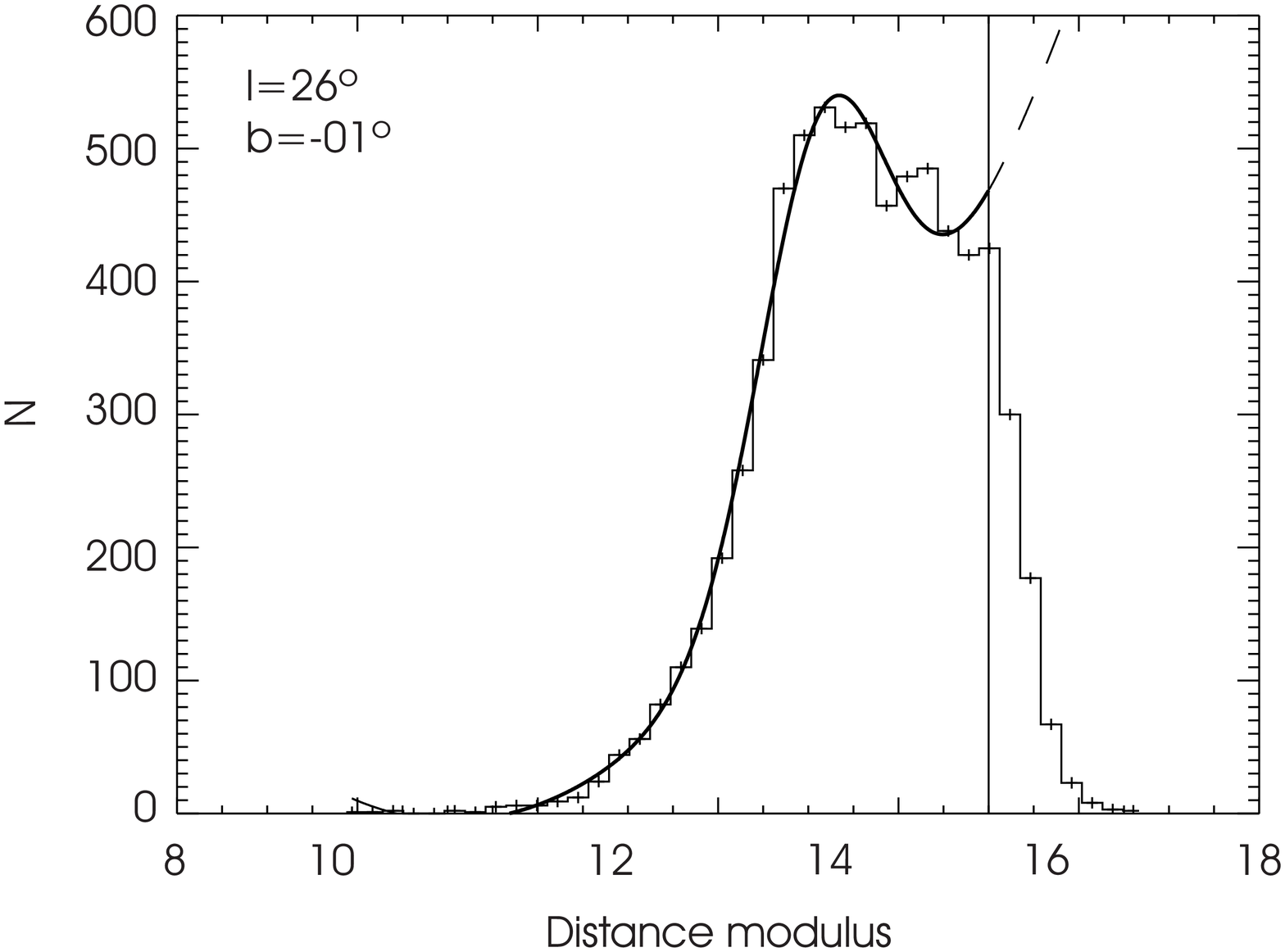}
\includegraphics[width=6.00cm]{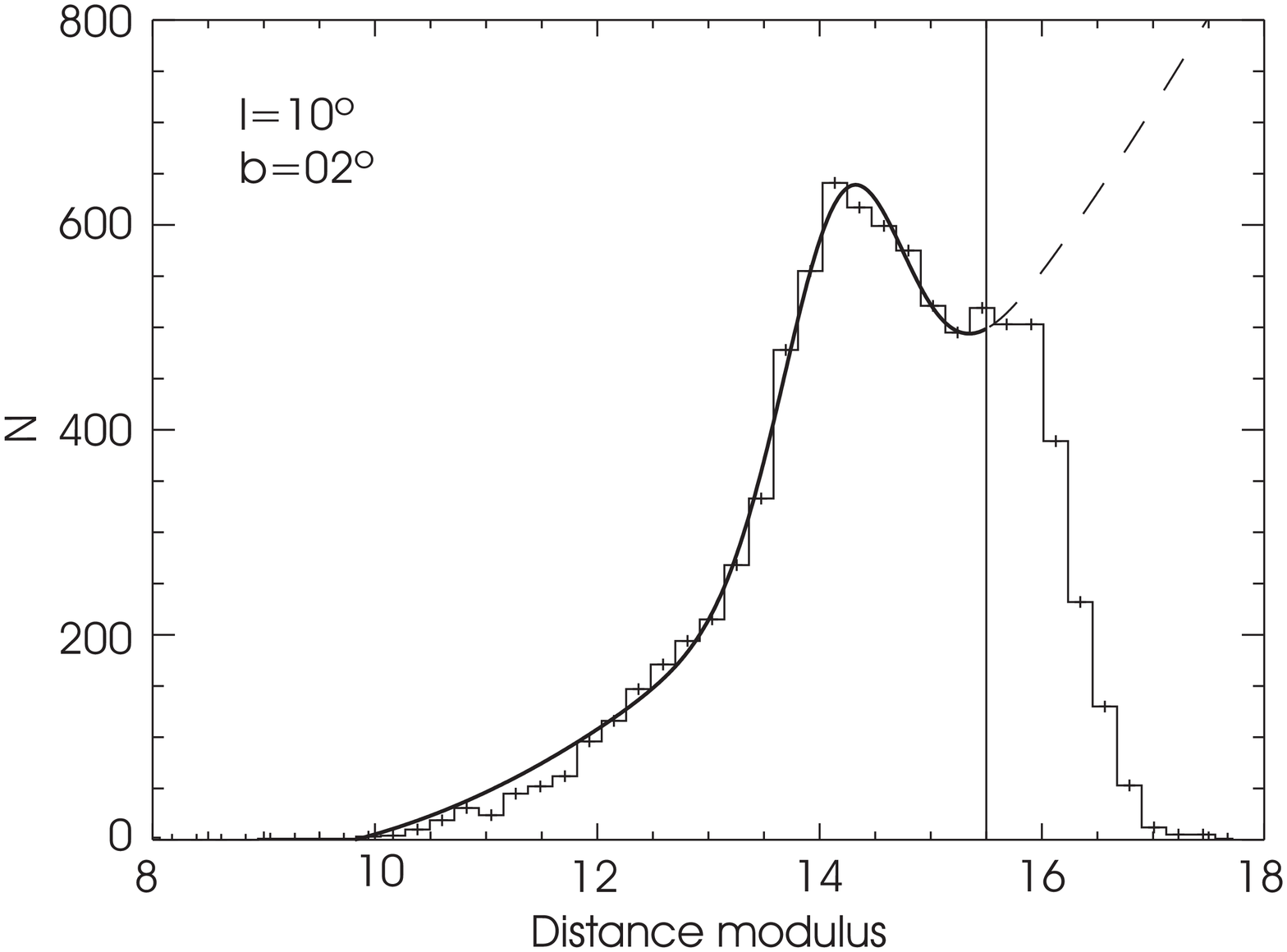}
\caption{Histograms of the distance modulus in three of the fields used in this work. 
Fits of eq.\ \ref{nm} to the histograms are shown as
solid curves, while vertical lines show the limiting magnitude estimates up to which the fits were performed.} 
\label{fits}
\end{figure}

The distances   to the   bar/bulge 
red-clump cluster of stars
are summarized in Table \ref{tabla1} and examples of the fits are shown in
Figure \ref{fits}. Error estimates for the distances
have been
obtained as a combination of the systematic uncertainties in  the
intrinsic colour
and absolute magnitude of the red-clump population and the fitting error
of eq.\ \ref{nm},
added in quadrature. The dispersion in distance is represented in each
case by $\sigma$. This is estimated  from a simple deconvolution:

\begin{equation}
\sigma = \sqrt{\sigma_{\rm RC}^2 - \sigma_0^2 - \sigma_e^2},
\end{equation}
 where $\sigma_{\rm RC}$ is the dispersion of the Gaussian fitted to the
histograms by
eq.\ \ref{nm}, $\sigma_0$ is the intrinsic dispersion of the red-clump
luminosity, which has been estimated in the range 0.15--0.2 mag (Alves 2000;
L\'opez-Corredoira et al.\ 2002) and finally, $\sigma_e$ is the photometric
error,  which for TCS-CAIN is of the order $\sim$0.08 mag
(Cabrera-Lavers et al.\ 2006).

\begin{table}[!h]
 \caption[]{Distance modulii and distances derived with the red-clump stars for
 the 49 fields used along this work, with each field covering 0.07 deg$^2$
 on the sky centred on the Galactic coordinates given in Cols. (1) and (2). The last eight fields are those in which $H$-band data were used to derive distances.}
\label{tabla1}
      \begin{center}
    \begin{tabular}{ccccc}
\hline
$l$ ($^\circ$) & $b$ ($^\circ$) & $\mu$ (mag) & $D$ (kpc) & $\sigma$ (kpc)\\ 
\hline
   18 &  0.0& 14.01$\pm$0.07&    6.317$\pm$0.229&      0.47 \\    
   20 &  0.0& 13.68$\pm$0.07&	 5.449$\pm$0.199&      0.44 \\    
   21 &  0.0& 14.05$\pm$0.09&	 6.455$\pm$0.298&      0.51 \\    
   22 &  0.0& 13.95$\pm$0.08&	 6.187$\pm$0.238&      0.47 \\    
   26 &  0.0& 13.81$\pm$0.07&	 5.778$\pm$0.186&      0.41 \\    
   27 &  0.0& 13.71$\pm$0.08&	 5.525$\pm$0.216&      0.42 \\    
   28 &  0.0& 13.66$\pm$0.11&	 5.392$\pm$0.303&      0.52 \\  
   29 &  0.0& 13.62$\pm$0.10&	 5.308$\pm$0.258&      0.50 \\  
   20 & $-$0.5& 13.36$\pm$0.08&	 4.704$\pm$0.184&	0.65\\  
   20 &  0.5& 13.79$\pm$0.08&	 5.719$\pm$0.221&	0.51\\  
   26 & $-$0.5& 13.58$\pm$0.08&	 5.211$\pm$0.204&	0.52\\  
   26 &  0.5& 13.65$\pm$0.12&	 5.366$\pm$0.304&	0.54\\  
   27 & $-$0.5& 13.55$\pm$0.08&	 5.126$\pm$0.202&	0.56\\  
   27 &  0.5& 13.72$\pm$0.10&	 5.536$\pm$0.248&	0.49\\  
   5 &  1.0& 14.25$\pm$0.07& 	 7.094$\pm$0.217&	0.70\\
   6 &  1.0& 14.55$\pm$0.08& 	 8.116$\pm$0.267&	0.42\\
   7 &  1.0& 14.15$\pm$0.08& 	 6.767$\pm$0.235&	0.67\\
   7 & $-$1.0& 14.06$\pm$0.07& 	 6.488$\pm$0.206&	0.45\\
   8 &  1.0& 14.38$\pm$0.09& 	 7.530$\pm$0.313&	0.71\\
   9 &  1.0& 13.97$\pm$0.09& 	 6.236$\pm$0.254&	0.62\\
   10 &  1.0& 14.29$\pm$0.10&	 7.197$\pm$0.332&	0.62\\
   15 &  1.0& 13.55$\pm$0.11&	 5.135$\pm$0.257&	0.45\\
   15 & $-$1.0& 13.33$\pm$0.09&	 4.626$\pm$0.188&	0.54\\
   20 &  1.0& 13.95$\pm$0.12&	 6.174$\pm$0.365&	0.47\\
   20 & $-$1.0& 13.72$\pm$0.13&	 5.555$\pm$0.335&	0.54\\
   23 &  1.0& 13.49$\pm$0.13&	 4.991$\pm$0.305&	0.58\\
   25 &  1.0& 13.86$\pm$0.11&	 5.906$\pm$0.322&	0.48\\
   25 & $-$1.0& 13.86$\pm$0.10&	 5.929$\pm$0.276&	0.49\\
   26 & $-$1.0& 13.51$\pm$0.12&	 5.028$\pm$0.327&	0.56\\
   3 & $-$2.0& 14.21$\pm$0.07& 	 6.943$\pm$0.229&	0.48\\
   5 & $-$2.0& 14.10$\pm$0.10& 	 6.603$\pm$0.298&	0.48\\
   7 &  2.0& 14.08$\pm$0.09& 	 6.542$\pm$0.264&	0.54\\
   7 & $-$2.0& 14.31$\pm$0.10& 	 7.277$\pm$0.333&	0.49\\
   10 &  2.0& 14.23$\pm$0.11&	 7.032$\pm$0.364&	0.50\\
   10 & $-$2.0& 14.36$\pm$0.12&	 7.444$\pm$0.421&	0.55\\
   17 & $-$2.0& 13.87$\pm$0.15&	 5.947$\pm$0.441&	1.01\\
   1 &  5.0& 14.57$\pm$0.12& 	 8.214$\pm$0.444&	0.45\\
   3 &  5.0& 14.28$\pm$0.12& 	 7.185$\pm$0.414&	0.46\\
   3 & $-$5.0& 14.35$\pm$0.12& 	 7.411$\pm$0.390&	0.47\\
   5 &  5.0& 14.16$\pm$0.12& 	 6.785$\pm$0.371&	0.43\\
   7 &  5.0& 14.04$\pm$0.09& 	 6.430$\pm$0.271&	0.72\\
\hline
  $-$5 &  5.0& 14.67$\pm$0.11&     8.588$\pm$0.459&	    0.43\\
  $-$3 &  5.0& 14.52$\pm$0.09&     8.015$\pm$0.340&	    0.49\\
  $-$2 &  5.0& 14.56$\pm$0.14&     8.183$\pm$0.502&	    0.48\\
   0 &  7.5& 14.35$\pm$0.13&     7.416$\pm$0.415&	    0.45\\
   5 & $-$3.5& 14.00$\pm$0.16&     6.319$\pm$0.431&	    0.42\\
   5 &  3.5& 13.72$\pm$0.14&     5.549$\pm$0.343&	    0.44\\
   5 &  7.5& 14.21$\pm$0.14&     6.943$\pm$0.524&	    0.50\\
   10 & 3.5& 13.84$\pm$0.16&     5.855$\pm$0.463&	    0.43\\

\hline
 \end{tabular}
\end{center}
\end{table}

\subsection{Contamination from the dwarfs}
Although  we have used the theoretical traces from the SKY model to isolate the red-clump population 
to avoid 
contamination due to other populations in the counts (Fig.\ \ref{trazas}), there will be some contamination from other 
sources and the effect of this on the overall shape of the reddening-corrected counts should be estimated.
For this, we have used the SKY model to predict  the relative
number of dwarf stars that will fall into the red-clump region of the CMDs (between the dashed lines in 
Figure \ref{trazas}). The dwarfs predicted by the SKY model that fell into the red-clump area on the CMD 
were subtracted from  the measured counts  to give a new ``corrected'' distribution. A full description of this
procedure can be found in section 3.3.3 of L\'opez-Corredoira et al.\ (2002). What is found is that for $m_K<12.5$ 
only  2.5--5\% of the detected sources are dwarfs, but that this rises to 10--40\% for $13<m_K<14$.

\begin{figure}[!h]
\centering
\includegraphics[width=7cm]{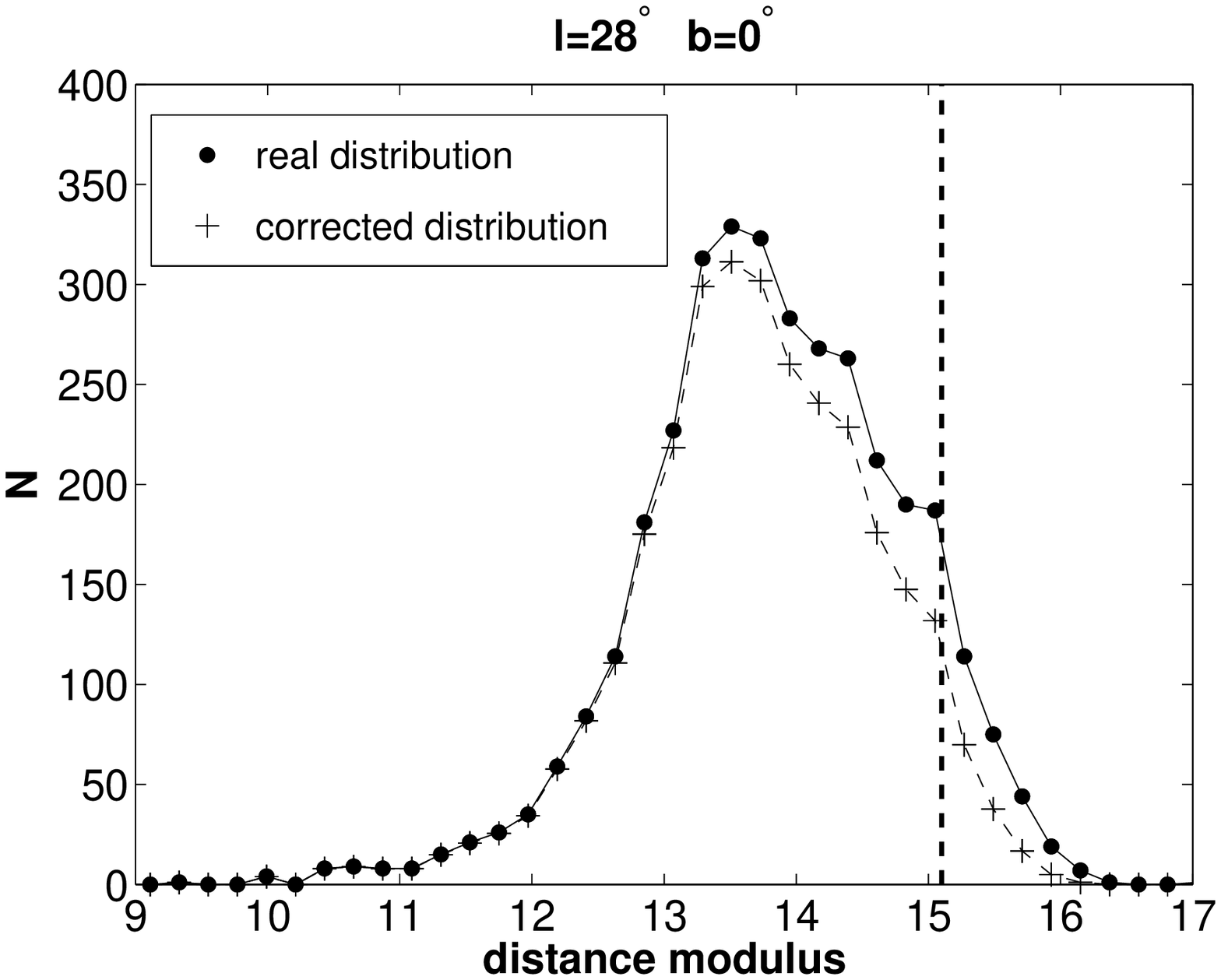}
\includegraphics[width=7cm]{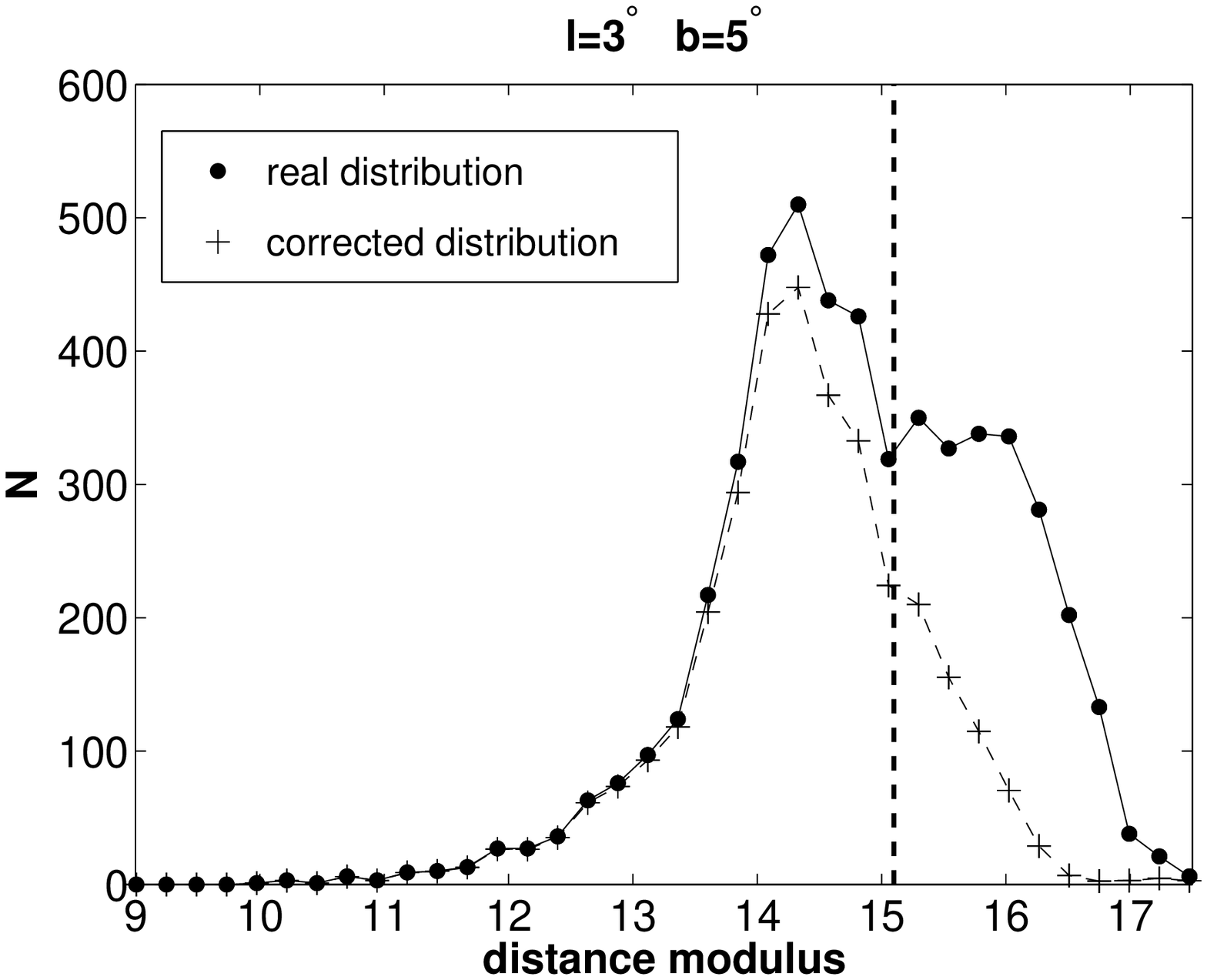}
\caption{Histograms of the distance modulus for two of the observed fields. Points/solid lines show the histograms 
derived directly from the
data, while dotted/dashed lines show the distributions once corrected for dwarf contamination (using the 
predictions of the SKY model).
Vertical dashed lines stand for the completeness-induced limits for each field.} 
\label{enanas}
\end{figure}

As shown in  Fig.\ \ref{enanas}, both for an
in-plane and an off-plane field, there is no significant difference in between the corrected and uncorrected
distribution. 
The effect of dwarfs is more important for fainter magnitudes
 (hence for apparently more distant sources), so the distributions overestimate the number of sources 
for higher values of the distance
modulus. However, the peak location is dominated by the red-clump population, and any possible contamination 
does not affect its location. We
fitted eq.\ \ref{nm} to the ``corrected'' distributions and obtained a mean difference in the distance 
modulus derived from either 
the uncorrected or
corrected data of 0.05 mag (which, translated into distances, yields to a difference of around 100 pc in this case),
which is  lower than the accuracy of the fit itself. This calculation was done only for a test sample of four randomly
selected fields. Since we did not find any significant difference in the results, correction was not attempted for
 with the remaining fields.

L\'opez-Corredoira et al.\ (2002) demonstrated that  the effect of
contamination is more severe off the plane, as extinction helps to separate the populations. 
The  results for off-plane fields show a
much better correction of the counts, giving a slightly narrower distribution around the peak. 
Although applying this  could in theory improve the fit, the effect is small and 
 introduces another factor based on models, hence we decided not to apply this correction.

\section{The Distribution of  sources in the inner Galaxy}

\subsection{In-plane fields ($|b|<1^\circ$)}
\label{secc:b0}
\begin{figure}[!h]
\centering
\resizebox*{5.22cm}{4.85cm}{\epsfig{file=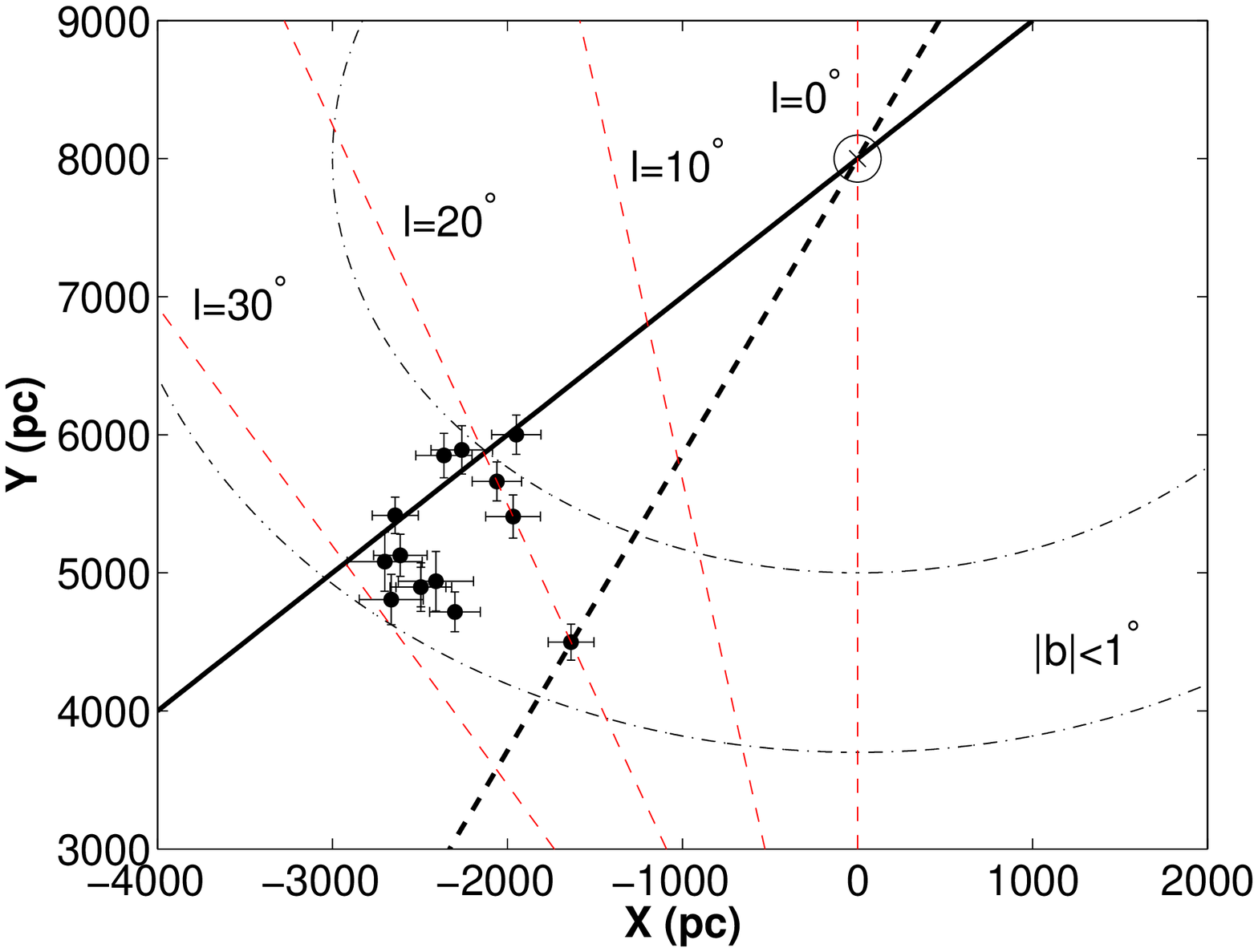}}
\resizebox*{5.22cm}{4.85cm}{\epsfig{file=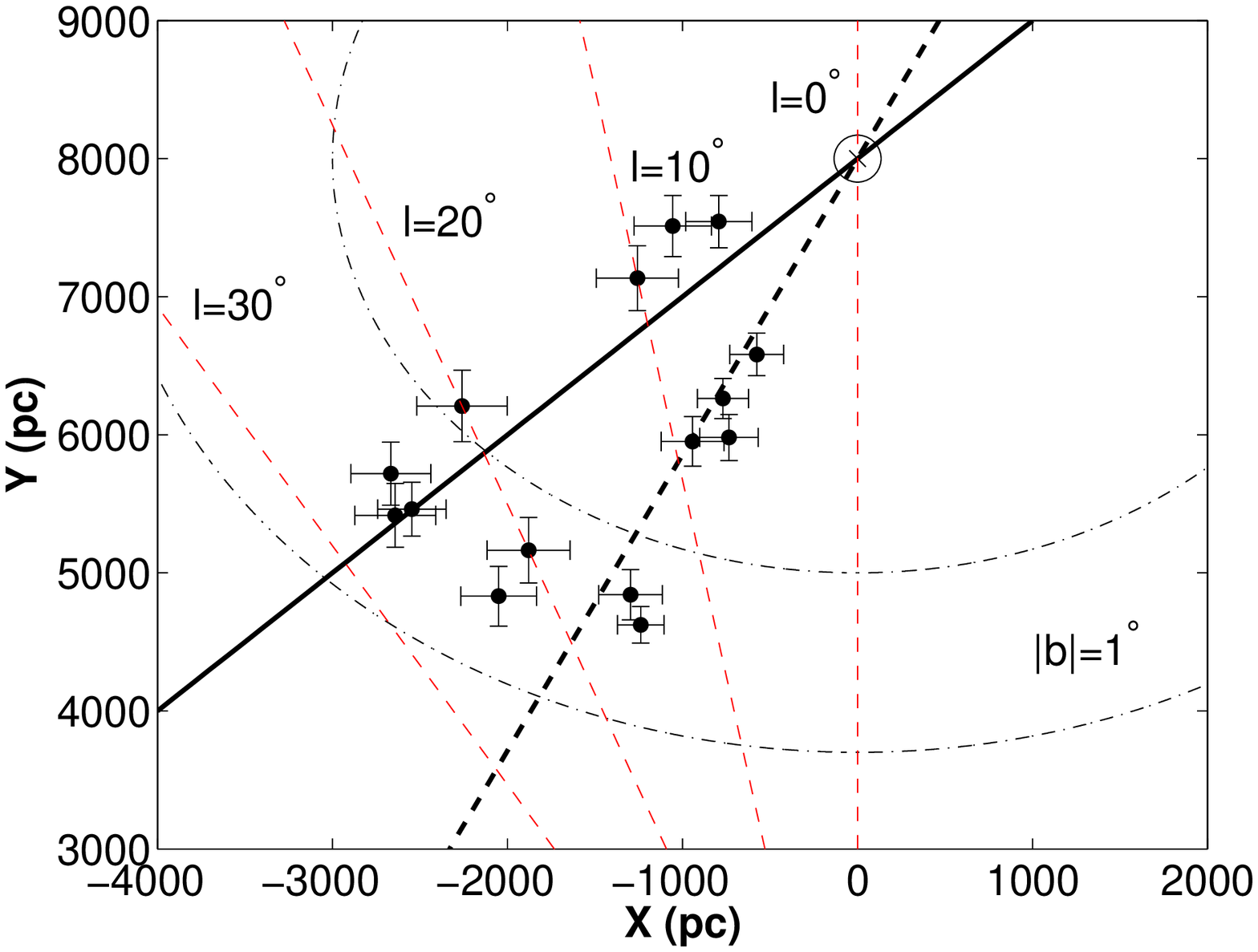}}
\resizebox*{5.22cm}{4.85cm}{\epsfig{file=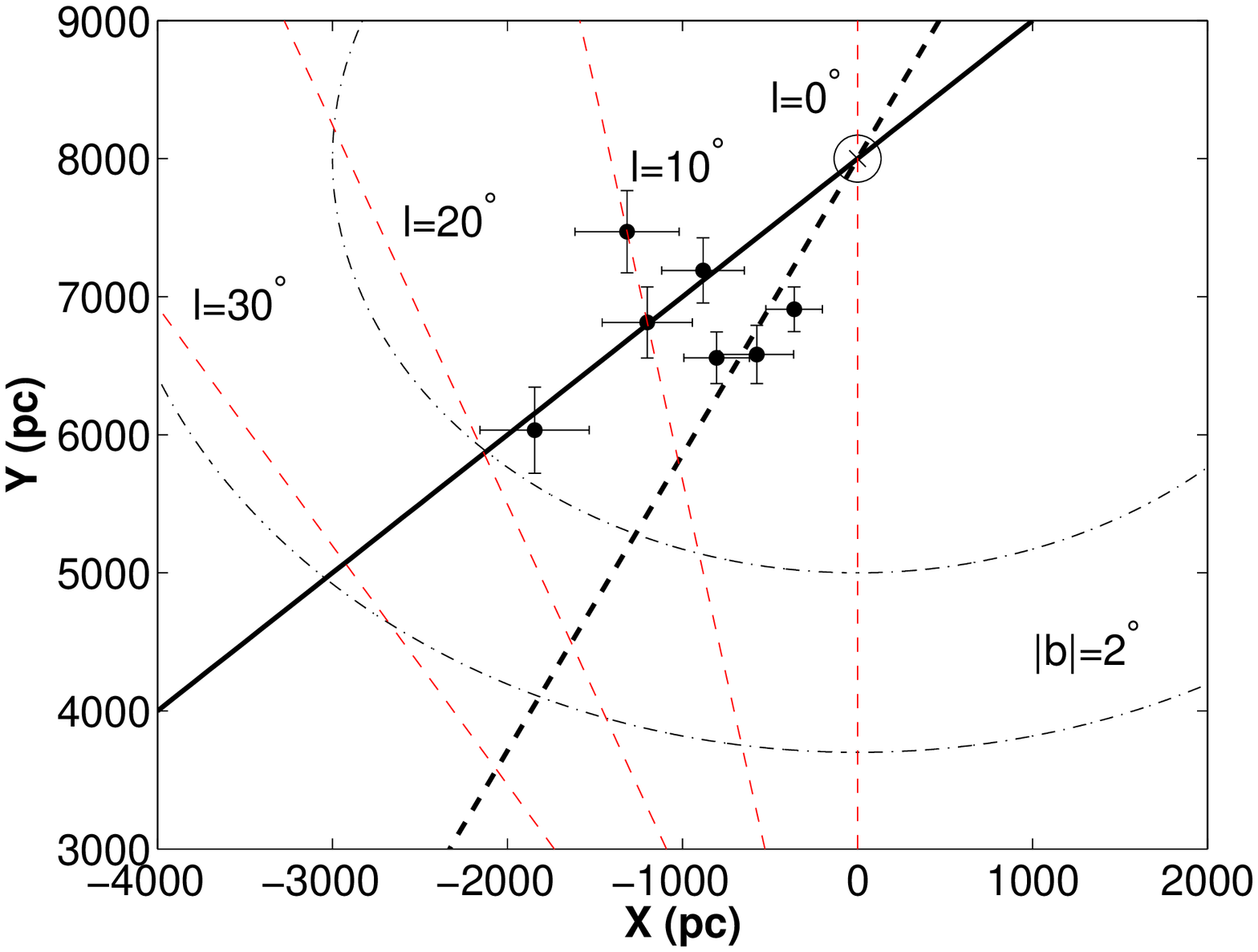}}
\resizebox*{5.22cm}{4.85cm}{\epsfig{file=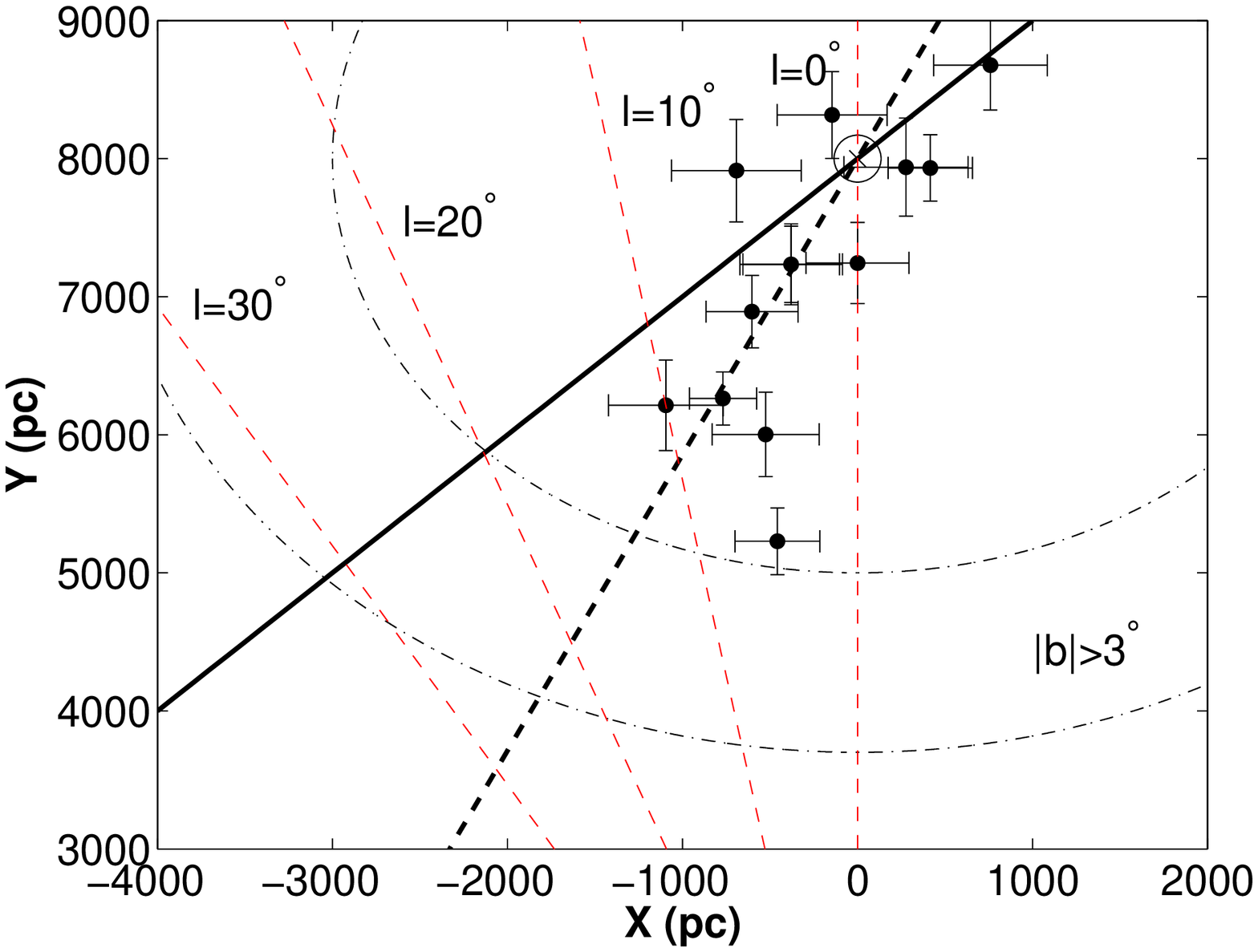}}
\caption{Spatial distribution of red-clump giants maxima in the 
$XY$-plane at different latitude intervals, with the Sun at (0,0) and the Galactic Center at (0,$8$ kpc) 
marked with a big circle. Two possible configurations for
the observed distribution are also shown: a feature with a position angle of
45$^\circ$ (solid line),  and another one with a
position angle of 25$^\circ$
(dashed line). Dot-dashed lines define two circles with radii 4.5 kpc and 3 kpc, respectively, while different lines of sight
towards the inner Galaxy are also shown in intervals of 10$^\circ$ in longitude. Error bars have been estimated from the distance
uncertainties, assuming these divided equally in both the $X$- and $Y$-axes.}
\label{bb}
\end{figure}

Results for in-plane fields in the range
 $18^\circ\le l\le 29^\circ$ show a  narrow distribution of distances from the
Sun. This is  
consistent with a high-density feature seen in the red-clump stars 
with position angle   40$\fdg$5 $\pm$ 3$\fdg$9
(after a linear fit through the points)
with respect to the
 Sun-Galactic Centre direction. First panel in Figure \ref{bb} shows the distribution of
red-clump density maxima for a face-on view of the Galaxy, with the
 Galactic Centre at
(0,8000 pc). The solid line shows a position angle of 45$^\circ$ (consistent with the 
43 degrees given for the bar in  Hammersley et al.\ 2000). The  dashed line
shows 
 a position angle of 25$^\circ$, which is consistent with the results of
Dwek
et al.\ (1995), L\'opez-Corredoira et al.\ (2005), Stanek et al.\ (1997) and,
 more recently, Babusiaux \& Gilmore (2005). The range of longitudes covered
in these fields ($18^\circ \le l \le 29^\circ$), however,  is very different 
from those which obtain the smaller position angle. Only a single point at
$l=20^\circ$, $b=-0.5^\circ$, is
observed near the latter position angle. This point 
  could be due to local highly anomalous extinction (see Sect. \ref{extc}), as this line of sight does run close to 
a major star formation region (Garay et al.\ 1998; Sridharan et al. 2005). 
 Furthermore, the over-density is is apparently somewhat 
inhomogeneous,  as was noted by Picaud et al.\ (2003) at  $l=20^\circ$ and $l=21^\circ$.

The gap in the longitude range $23^\circ\leq l \leq 25^\circ$  is due to the
very high extinction that
extends up to $|b|\sim0.4^\circ$, which is noticeable also in the mid-infrared
range (see Fig.\ 1 of Benjamin et al.\ 2005). This makes it  impossible to obtain
an accurate fit for the fields located in this region. The same problem applies
to the in-plane data for $0^\circ\leq l \le 15^\circ$. In these fields the 
completeness limit is far brighter than the apparent magnitude of the K giants for any structure that would be 
present, that is significantly fainter than the completeness limit. Hence, the above method cannot be applied.
However, in the CMD the giant branch from the feature is still visible but its distance cannot be measured.

\subsection{Off-plane fields ($|b|>1^\circ$)}

\begin{figure}[!h]
\centering
\resizebox*{7cm}{6.5cm}{\epsfig{file=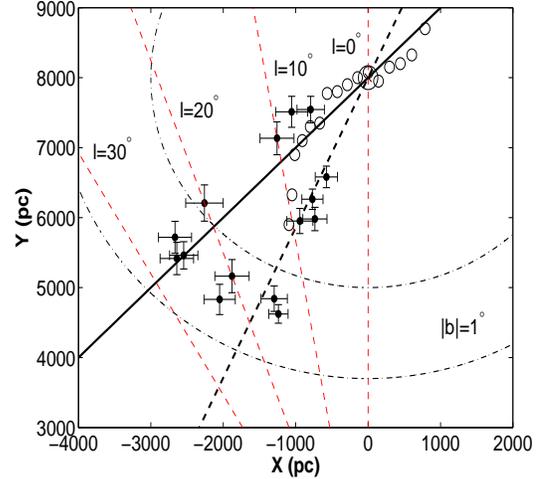}}
\caption{Spatial distribution of red-clump giants maxima presented face on 
for data at $|b|=1^\circ$ and the results obtained for Nishiyama et al.\
(2005) for comparison (shown as open circles).}
\label{nishi}
\end{figure}

The results for  $|b|=1^\circ$ (second panel in Fig.\
\ref{bb}) show that there are two distance regimes
 for the red-clump population, one that follows approximately  the
 43$\fdg$4 $\pm$ 3$\fdg$5 angle 
  (nine
 points) and a second with a smaller angle of
21$\fdg$9 $\pm$ 1$\fdg$0  (six points). Most of the points in this latter
feature are at  $l<15^\circ$, while the points associated with the former are
mostly located at $l\ge20^\circ$. Nishiyama et al.\ (2005) also
observed the red-clump stars  at $|b|=1^\circ$ over the range 
$8^\circ\le l \le 15^\circ$ and there is good agreement between 
their distribution of red-clump maxima and ours (Fig.\ \ref{nishi}), even in the central regions of the Galaxy.

For  $|b|=2^\circ$ the results are less conclusive (third panel in Fig.\ \ref{bb}). 
This is because we are not deep enough to be complete in the red-clump giants
 in the innermost Galaxy ($|l|<5^\circ$) and this makes it difficult to observe
the peak in the red-clump counts. Also, for fields
$|l|>15^\circ$ there is no red-clump cluster observed in the modulus distance
histograms, i.e.\ there is no feature there. Only one point is detected  with
$l>15^\circ$ and this has  the highest dispersion in distance of any field.
The remainder of the fields lie where there is little difference in
distance   between the two position angles, making it difficult 
to clearly assign the red-clump density maxima to one of them.  A linear fit to 
those data yields a global position angle of  
36$\fdg$6 $\pm$ 5$\fdg$8
 (after discarding those at $l=17^\circ$), although if the points are assigned
 to the position angle that runs closer to that position, and are then averaged separately  
 we obtain
 position angles of 44$\fdg$1 $\pm$ 2$\fdg$8  and
25$\fdg$1 $\pm$ 2$\fdg$7, respectively.

Farther away from the plane ($|b|>3^\circ$) it is possible to 
sample the innermost Galaxy as the incompleteness effect disappears; however,
there are no observable features in the histograms for $l>7^\circ$ as
the disc component dominates and there is no dense structure at a specific distance. 
The distribution of red-clump peaks
shows in this case a position angle of 30$\fdg$7 $\pm$ 5$\fdg$2 with
respect to the Sun--Galactic Centre line
(lower panel in Fig.\ \ref{bb}). 
These results are compatible
with the position angle of 29$^\circ$ $\pm$ 8$^\circ$ derived in
L\'opez-Corredoira et al.\ (2005) for the  bulge of the Milky Way 
using 2MASS
data. Also, this result is consistent with the position angle of
22$^\circ$$\pm$5$\fdg$5 recently derived
by Babusiaux \& Gilmore analysing their private in-plane data at
$|l|<10^\circ$.

\begin{table}[!h]
 \caption[]{Fitted position angles and mean dispersion in distances for
the different latitude ranges.}
\label{tabla2}
      \begin{center}
    \begin{tabular}{ccccc}
\hline
 $b$ range & $N$ & $\phi_{\rm bar}$ ($^\circ$) & $\phi_{\rm bulge}$ ($^\circ$) &
$\langle\sigma\rangle$ (kpc)\\
\hline
\hline
$|b|<1^\circ$ & 14 & 40.5$\pm$3.9 & -- & 0.48$\pm$0.04\\
$|b|=1^\circ$ & 15 & 43.4$\pm$3.5 & 21.9$\pm$1.0 & 0.55$\pm$0.09\\
$|b|=2^\circ$ & 7 & 44.1$\pm$2.8 & 25.1$\pm$2.7 & 0.51$\pm$0.03\\
$|b|\ge3.5^\circ$ &13 & -- & 30.7$\pm$5.2 & 0.45$\pm$0.02\\
$|b|=3.5^\circ$ &3&  -- & 21.6$\pm$3.9& --\\
$|b|=5^\circ$ & 8 & -- & 31.5$\pm$3.5 & --\\
\hline
 \end{tabular}
\end{center}
\end{table}

\subsection{The  effect of extinction on the results}
\label{extc}
A global extinction law is assumed when  deriving the reddening-independent
magnitudes. However,  recent
results show that there are noticeable differences in the
extinction law derived in different directions towards the inner Galaxy
(Nishiyama et al.\ 2006a) and even very small variations in the ratio of
total to selective extinction are present on very small scales (Gosling
et al.\ 2006), so care must be taken  when applying reddening
corrections. The combination of this plus the possible mixture of
sources coming from the two position angles  would broaden
in the distance histograms, hence giving a higher dispersion in
distances, $\sigma$.

\begin{figure}[!h]
\centering
\resizebox*{5.33cm}{5.05cm}{\epsfig{file=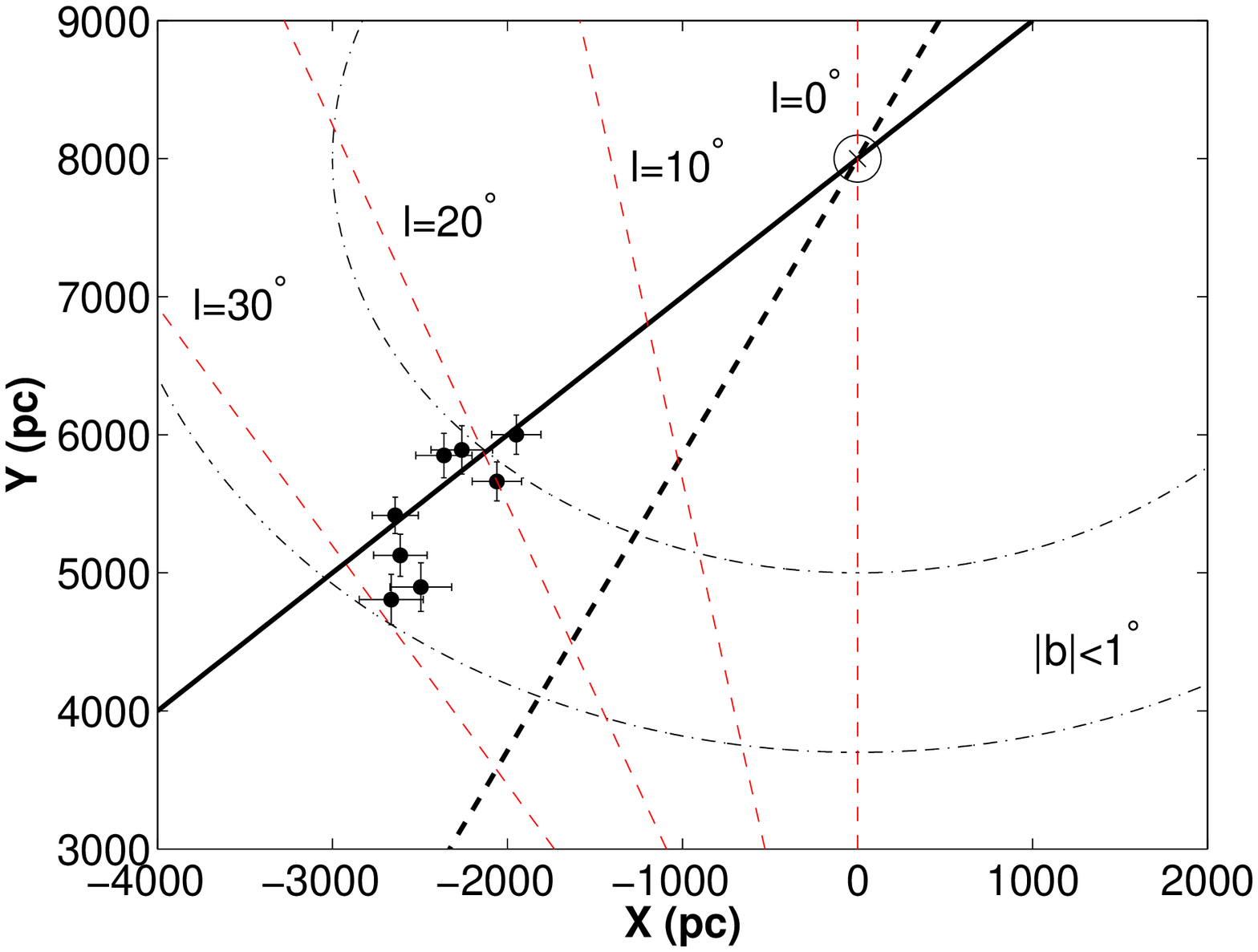}}
\resizebox*{5.33cm}{5.05cm}{\epsfig{file=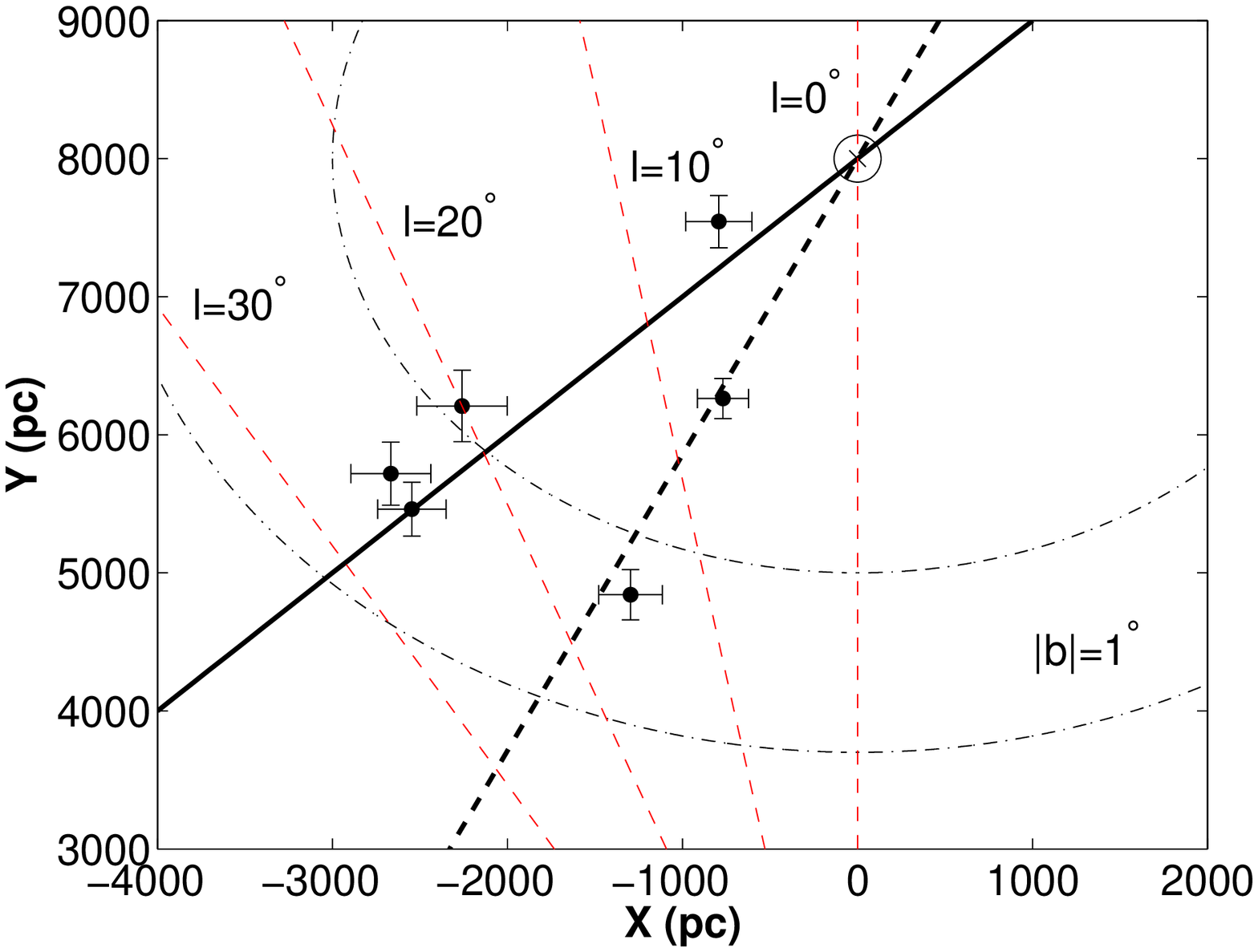}}
\resizebox*{5.33cm}{5.05cm}{\epsfig{file=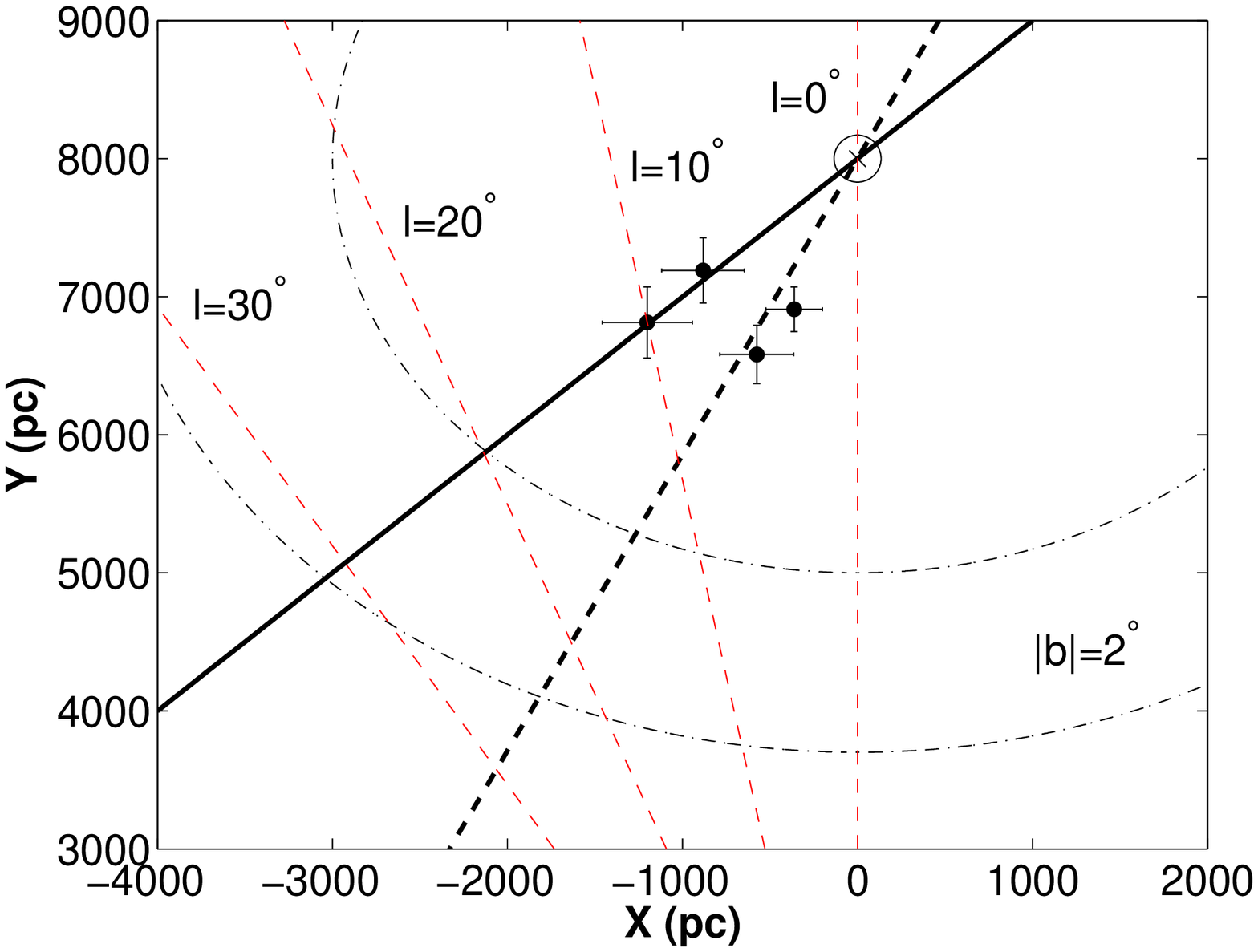}}
\resizebox*{5.33cm}{5.05cm}{\epsfig{file=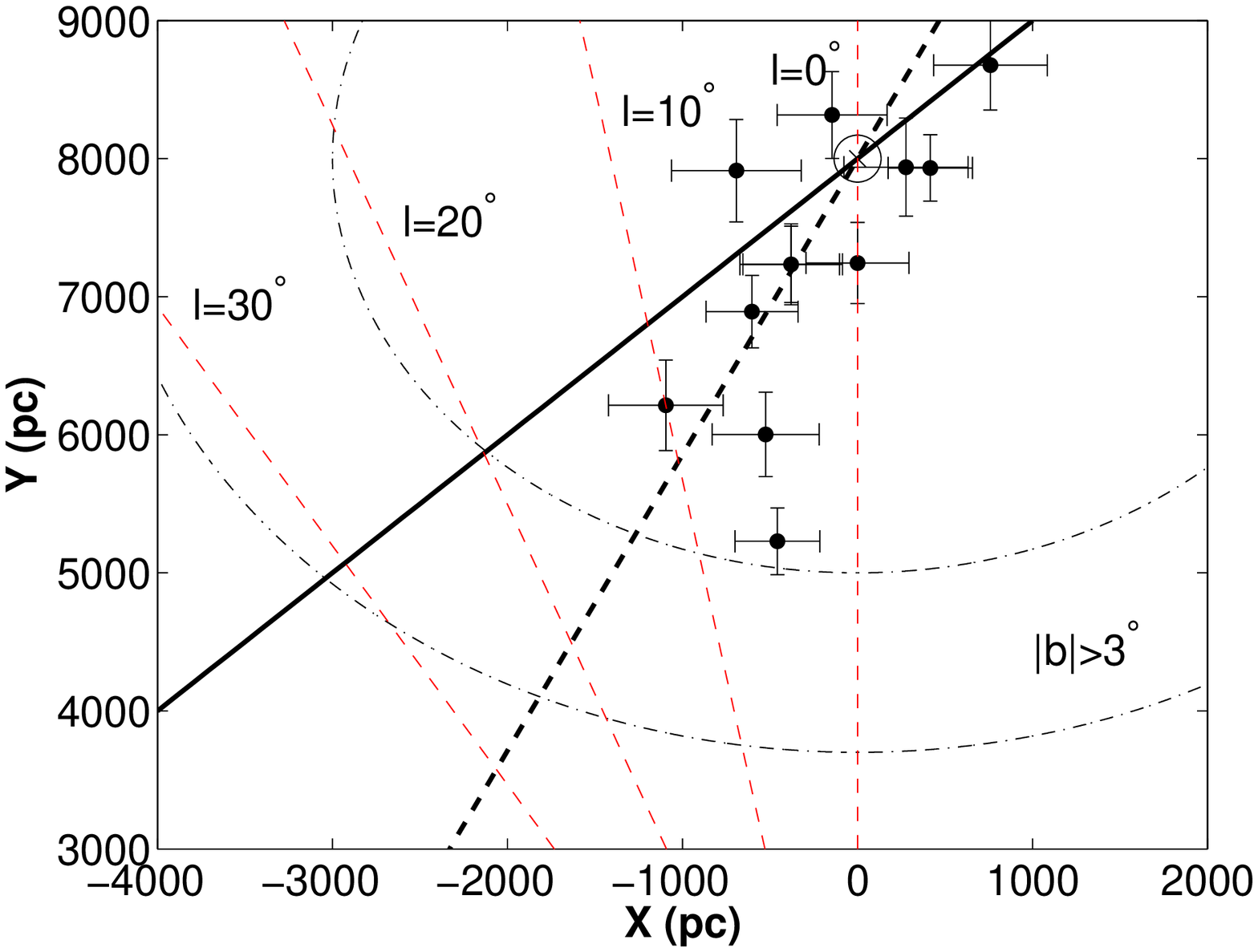}}
\caption{Same as in Fig.\ \ref{bb}, but considering only those fields
where dispersion in distances is less than 0.5 kpc.}
\label{bbsigma}
\end{figure}

To examine  this effect, we have calculated the
mean dispersion in distances in
 the four latitude ranges showed in Fig.\
\ref{bb} (Table \ref{tabla2}).  There is clearly less
dispersion  at $|b|<1^\circ$ and at $|b|>3^\circ$ than at
intermediate latitudes, where the mixture of sources becomes more
important (error bars shown in Table \ref{tabla2} account only for the
standard deviation of values). The mean values all cluster around 0.5 kpc,
but again, with slightly
larger values at $|b|=1^\circ$ and $|b|=2^\circ$. Hence, if we consider the results only for those fields
that present $\sigma \le 0.5$ kpc we obtain the red-clump peak
distributions shown in Figure \ref{bbsigma}.

It is noticeable that in all the cases those points that were
originally located  at positions intermediate  between the two position angles 
have been removed. Furthermore, only one
point has disappeared from the fields  at $|b|<1^\circ$ and $|b|>3^\circ$ as a
reflection of the lower dispersion values obtained in those latitude
ranges. The point removed is that noted in \S\ref{secc:b0} at $l=20^\circ$, $b=-0.5^\circ$
 suggesting that   extinction is
responsible for its discrepant  location.

\subsection{Discussion of the two position angles}

The above shows that:
\begin{itemize}
\item The red-clump stars show that there is a dense feature  seen on the plane 
in lines of sight $0^\circ<l<28^\circ$,  but at $|b|>2^\circ$ it is only seen  for  
$0^\circ<l<10^\circ$. In other locations there is 
no dense feature, and only the disc is seen. 
\item Away from the Galactic plane for $|l|<10^\circ$ the peaks in density of  
the red-clump population follow a very well defined feature with a position angle of  
23$\fdg$1 $\pm$ 0$\fdg$9 with respect to the Sun--Galactic Centre line.
\item  In the plane ($|b|<1^\circ$) for $|l|>18^\circ$ there is a longer structure with a distinct
position angle of   43$\fdg$0 $\pm$ 1$\fdg$9. This feature is not seen more than 2 degrees away
from the plane. 
\item At intermediate latitudes, there
is a mixture of sources coming from
 both structures producing a more spread out distribution in the
distances derived from the red-clump location in the CMDs. 

\end{itemize}
The position angles are derived by a weighted mean of the values summarized in
Table \ref{tabla2} for both features. 

Sevenster et al.\ (1999) found a similar effect when comparing $N$-body models with their 
sample of OH/IR stars and noted that the position angle changed from  $\phi$ = 44$^\circ$ when
 using in-plane fields to $\phi$ = 25$^\circ$ when only high latitude
fields were considered. They also noted in the literature that lower values for the viewing angles determined from stellar data 
 were found when low-latitude fields were excluded from
the fit, which is in agreement with the work presented here.

The feature seen on the plane $18^\circ<l<27^\circ$  in the red-clump stars can have 
no direct link with the Scutum spiral arm. The tangential point to the stars in this arm is at 
$l\sim 32^\circ$, and it would be expected that any features associated with an arm would be strongest there. 
However, there is no red-clump cluster seen at that location 
(Hammersley et al.\ 2000), and inwards of the this  point the effect of the arm would
become less (Hammersley et al.\ 1994). Furthermore, a spiral arm would not  form a
feature that appeared to run directly towards the Galactic Centre. 

\begin{figure}[!h]
\centering
\resizebox*{7cm}{6.5cm}{\epsfig{file=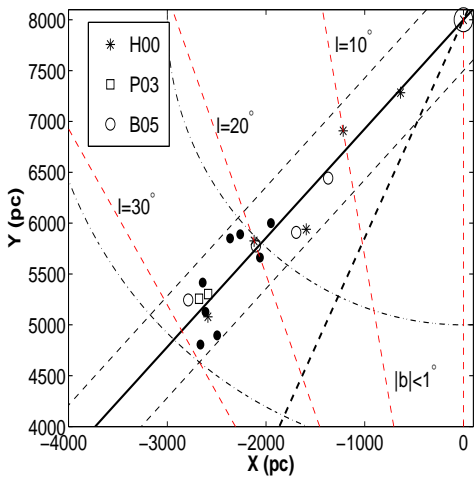}}
\caption{Spatial distribution of red-clump giants maxima for in-plane fields in the $XY$-plane, with some of the more 
recent estimates for this feature (H00: Hammersley et al. 2000; P03: Picaud et al. 2003; B05: Benjamin et al. 2005). Note there is an
excellent agreement between the different sets of data. The long bar with a position angle of 43$\fdg$0 derived in this work is 
represented by a solid line (and a width of 1 kpc is represented by dashed lines).}
\label{b0_1}
\end{figure}

 Extinction can also be discounted.  There are a range of extinctions covered
 between 18$^\circ \le l \le 29^\circ$ and in no field is the extinction less 
than is predicted by the simple  extinction model used in Wainscoat et al.\ (1992). In any case, 
the method used corrects for extinction. Furthermore, our result is in extremely
good agreement with the result of   Benjamin et al.\ (2005)  (see Fig.\ \ref{b0_1}), where the effects of extinction are
significantly reduced. Also, in Fig. \ref{b0_1}
we show how the concentration of data
seen at $18^\circ<l<27^\circ$ is not an isolated
feature, and how it is consistent with other estimates for this 
structure obtained with
deeper data than ours in different lines of sight to 
the inner Galaxy. In fact, it is
clear that this feature runs to the inner Galaxy, at least up to 1.5-2 kpc from the
Galactic Centre, where it merges with the bulge.

The only scenario that readily fits the star count data and CMDs, those presented both here and elsewhere,   
is that of  a triaxial bulge plus a long  bar. The bulge has a position angle of about 25$^\circ$ and dominates 
the counts to $|l|\sim12^\circ$ 
and up to $|b| =7^\circ$ or farther off the plane. The in-plane bar has a position angle of 43$^\circ$ with a 
half-length of 
some 4~kpc but is only detected close to the plane. This bulge + bar picture is not so unreasonable.
Recent $N$-body simulations on the secular evolution of disc galaxies predict that in
a bar-driven evolution, the bar has a 
thick inner part of shorter extent and a thin outer
part of larger extent (Athanassoula 2006), a fact that is in good 
agreement with observations (Athanassoula 2005; Bureau et al.\ 2006). Also, 
the ratio of the thin bar
size to the inner bulge length that is derived from our data (between 2.2--2.8 assuming a
bulge length of 1.5--2 kpc) is in completely agreement with that was 
observed in external galaxies (2.7 $\pm$ 0.3, L$\ddot{u}$tticke et al.\ 2000).

\begin{figure}[!h]
\centering
\resizebox*{7cm}{6.5cm}{\epsfig{file=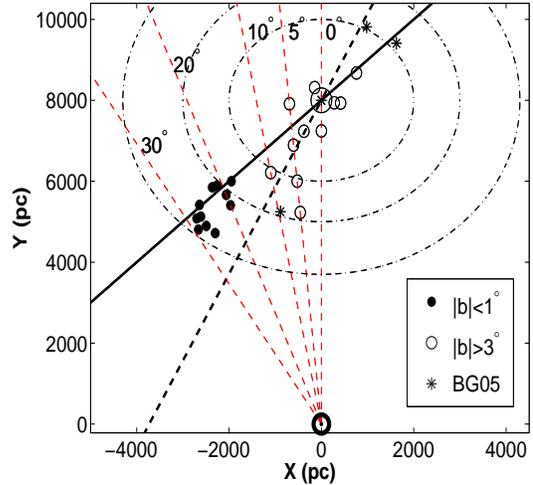}}
\caption{Distribution of positions for the red-clump sources obtained in
this work at $|b|<1^\circ$ and $|b|>3^\circ$ compared with the
results of Babusiaux \& Gilmore (2005) by using in-plane NIR data. 
 Dot dashed lines delimite circles with radii 4.3 kpc, 3 kpc and
2 kpc, respectively.}
\label{bbab}
\end{figure}

In order to illustrate this, Fig.\ \ref{bbab}  shows the results from Babusiaux \& Gilmore (2005; hereafter BG05).
Also shown on the plot are the results from this work and position angles of 45$^\circ$ and 25$^\circ$ degrees. 
The BG05 data set is deeper than ours (as a result of a  pixel scale five times smaller),
permitting them to reach the inner Galaxy with in-plane data.  It is noticeable that  the BG05 points at  
$l\sim 10^\circ$ and $l\sim -5^\circ$ coincide with positions for the triaxial bulge obtained here; however, the point 
at  $l\sim -10^\circ$ lies on the 45$^\circ$ position angle and is well away from the position required for the
triaxial bulge.

\subsection{Scale height of bar/bulge sources}
\begin{figure}[!h]
\centering
\includegraphics[width=8cm]{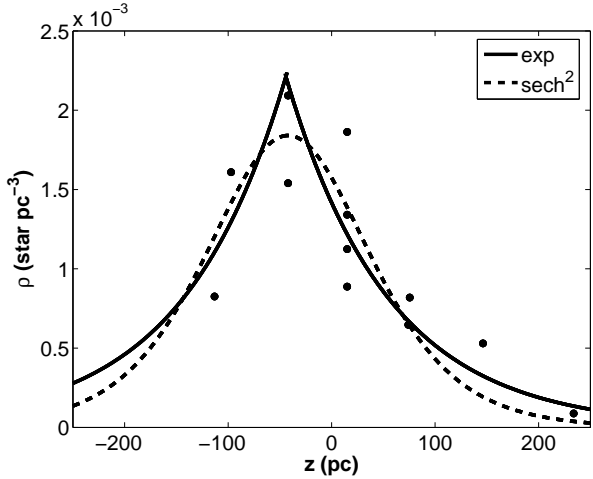}
\caption{Variation of the space density  as a function of distance from the Galactic plane. For 
comparison, the best fit of an
exponential law (solid line) and a sech$^2$ law (dashed line) are shown.} 
\label{escala1}
\end{figure}

\begin{figure}[!h]
\centering
\includegraphics[width=8cm]{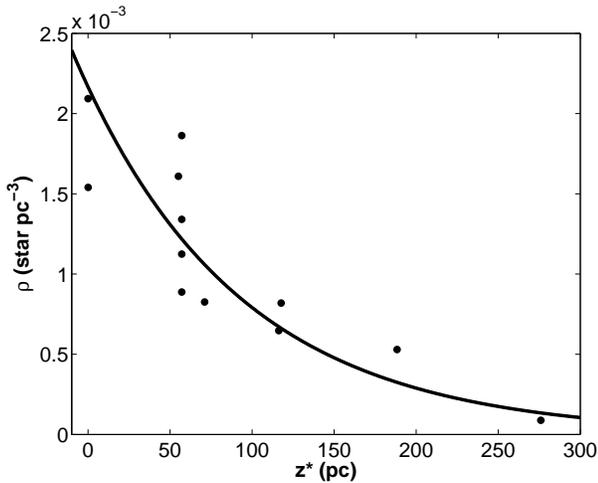}
\caption{Vertical dependence of the density of bar stars. The best exponential fit, with a scale 
height of 99.28 $\pm$ 3.13 pc, is also shown as a solid line.} 
\label{escala}
\end{figure}

From the histograms of the red-clump  sources an estimate can be made of the mean scale height
of the sources associated with the long thin bar. We have used only those
fields with $l\ge 20^\circ$ and $|b|\le1^\circ$, as 
 at $b=2^\circ$ there are not sufficient bar sources for a reliable extraction. 
For each field the space densities along different lines of sight were determined by selecting the stars within
one $\sigma$ of the maximum of the histograms of the distance modulus (see Table
\ref{tabla1}).
Combining  the distances along the line of sight 
to the features ($d$) and the Galactic coordinates of the fields ($l,b$) allows these space densities to be expressed with 
 respect to the galactocentric distance  ($R$) and the height ($z$) above the plane, taking into
account the location of the Sun 15~pc above the mean plane (Hammersley et al.\ 1995).
 The result is
shown in Figure \ref{escala1}. It can be observed that the mid-point of the distribution is at $z=-41$ pc.  Hence, we have 
represented in Figure \ref{escala} the space density with respect to a vertical distance to the 
plane $z^*$, defined as $z^*=|z+41|$. This means that the location of this hypothetical long thin bar 
is $\sim 0.5^\circ$  tilted with respect to the mean plane, a result that is compatible
with the values found in Picaud \& Robin (2004) for the orientation of the outer bulge.

We have also tested a sech$^2$ profile to compare with the observed space density for a simple exponential. 
The two equations used are: 
\begin{center}
$\rho=\rho_0 \exp(-z^*/h_z)$,\\
$\rho=\rho_0 sech^2(z^*/h_z)$
\end{center}
where $h_z$ is the scale height of the bar sources and $\rho_0$ is the space density in the Galactic plane. 
The exponential gives $h_z$ = 99.3 $\pm$ 3.1 pc and $\rho_0$ = 0.022 $\pm$ 0.014 pc$^{-3}$, whereas the 
sech$^2$ profile yields
$h_z$ = 105.4 $\pm$ 5.3 pc and
$\rho_0$ = 0.018 $\pm$ 0.013  pc$^{-3}$. Both fits are shown in Figure \ref{escala1}.

\begin{figure}[!h]
\centering
\includegraphics[width=8cm]{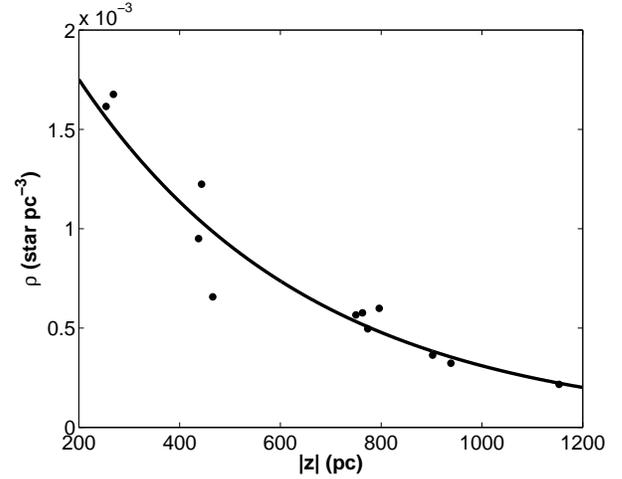}
\caption{Vertical dependence of the density of inner bulge stars. The best exponential fit shown as a solid line now  has 
a scale height $\sim$5
times higher than that of the thin bar.} 
\label{escalabulbo}
\end{figure}

An exponential scale height for the bar of  around 100 pc is a factor 2 higher than that  previously
obtained by L\'opez-Corredoira et al.\ (2001) or Hammersley et al.\ (1995) from NIR star counts.
 However, it should be noted that here we are looking at a very different star type than in the previous studies. 
 In both of those cases the stars observed were  considerably more luminous, and spectra of a selection of these sources 
 (Garz\'on et al.\ 1997)
  showed that there were a large number of young sources at $l=27^\circ$ and $l=21^\circ$. 
  The red-clump sources observed here probably  represent 
an older population and hence the larger scale height would be expected. However, the  scale height for the red-clump 
stars in other
Galactic components is far higher. For the thin disc it is $\sim$260 pc  (Cabrera-Lavers et al.\ 2005)
 and applying the same method used above  to the 
 bulge for the fields at  $|b|\ge3.5^\circ$ gives a  scale height of  
 $h_z=488\pm 28$ pc (see Fig. \ref{escalabulbo}).  Hence the scale height is a further clear indication that on the 
plane between
 $18^\circ\le l\le 29^\circ$ there is a geometric component that is very different  to the bulge or thin disc.

\section{Systematic uncertainties in the red-clump method}

\subsection{Deriving distances from the number count histograms}
The distances to the red-clump population were obtained by fitting  eq.\ \ref{nm} to the
distance modulus histograms. However, the maximum of star counts vs. magnitude is not strictly coincident with
the maximum in density along the line of sight, which is proportional to the
first multiplied by a factor 1/$d^3$. Hence the magnitude histogram counts, $N(m)$,  should be transformed into 
density along the line of sight, $\rho(d)$,
 and eq.\ \ref{nm} then fitted to the resultant distribution. From the relationship between $\rho(d)$ and $N(m)$ 
it can be shown 
 that the difference between the corrected distance to the maximum in the density
distribution ($r^*_m$) and the distance obtained to the maximum in the counts histograms ($r_m$) is:
\begin{equation}
r^*_m = r_m + \Delta r_m
\end{equation}
\begin{equation}
\Delta r_m = \frac {3 ~\rho (r_m)}{r_m \rho''(r_m)}
\end{equation}
$r_m$ being the distance to the maximum in the $N(m)$ histogram (hence, $N'(m[r_m])=0$) and $\rho''(r_m)$ the 
second derivative of $\rho$ with
respect to the distance, measured at $r_m$. As $\rho''(r_m)<0$, then $r^*_m < r_m $, thus the corrected distances are
 slightly lower than those
obtained from the maximum of the magnitude histograms. 

\begin{figure}[!h]
\centering
\includegraphics[width=7cm]{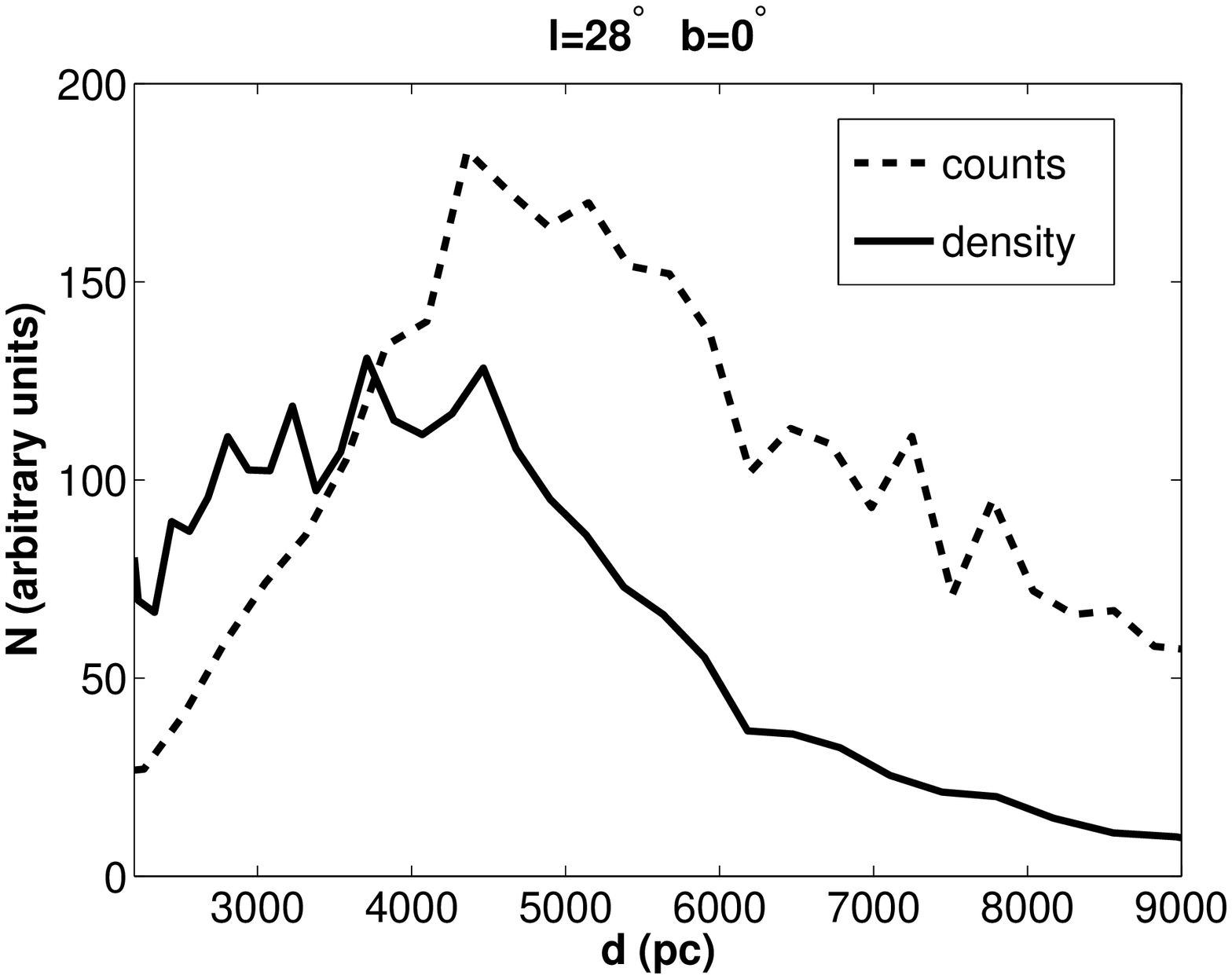}
\includegraphics[width=7cm]{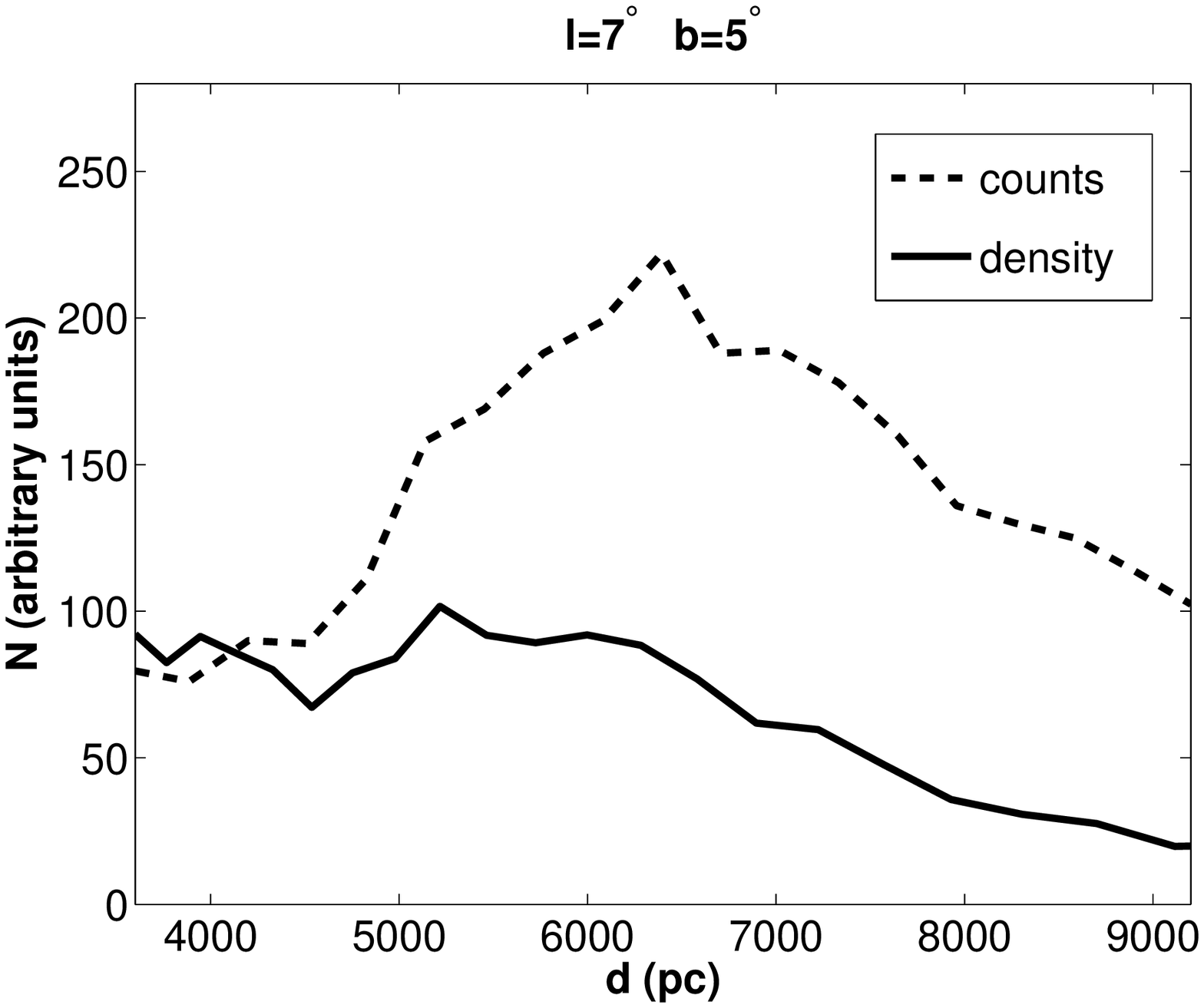}
\caption{Comparison between the density (solid line) and the number count (dashed line) distributions with
 respect to the distance along the
line of sight for long bar (above) and  bulge (below) fields as examples. Density distributions are arbitrarily 
displaced along  the $Y$-axis in order to
make both distributions comparable.} 
\label{delta}
\end{figure}

This effect is more noticeable for the bulge component than for the long thin bar, as $\rho''(r_m)$ is very 
large for this latter. We have 
estimated
the range of values for $\Delta r_m$  for the bulge and for the long bar, obtaining the true density distributions 
along the line of sight. We
have used those fields with $|b|\ge3.5^\circ$ as representative of the bulge, while those in-plane fields with 
$l>20^\circ$ are representative of
the long bar. It is found that $\Delta r_m$ is in the range  100--300 pc in bulge fields, whereas for long bar fields the effect 
is far smaller, in the order of 25--50 pc. Figure \ref{delta} shows two examples of the change in the overall shape 
of the density
distribution compared to the number count histograms used to derive the distances to the red-clump population. 
The effect on the determined geometries for either the bulge and the long bar is not large in either case. 
For the long bar it is  ten times lower than the 
uncertainties in the fit itself (see Table \ref{tabla1}) and  even for the bulge it is only  about 50\% of the
uncertainty in the fit.

\subsection{Difference in maximum density along the line of
sight and the position of the major axis of a triaxial structure}

\begin{figure}[!h]
\begin{center}
\vspace{1cm}
\mbox{\epsfig{file=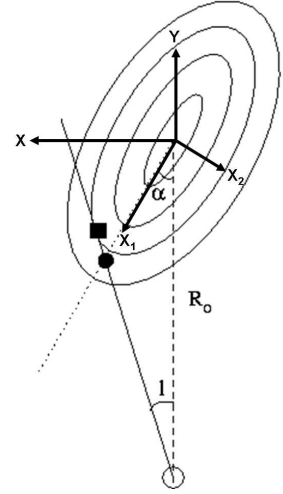,height=6.5cm}}
\end{center}
\caption{Schematic representation of the difference between the
maximum density position along the line of sight (filled square) for
triaxial
ellipsoids and the intersection of their major axis with the line of
sight (filled circle).}
\label{Fig:eliptan}
\end{figure}

The position of the maximum density
along the line of sight is not coincident with the
the position of the major axis of a triaxial structure unless
the structure  is very thin along the line of sight (a bar). There are two
effects:
\begin{itemize}
\item The more  important one
is that the density along the line of sight reaches a maximum at the tangential
point of the innermost  ellipsoid along it and this is not in general
on the major axis (see Fig. \ref{Fig:eliptan}).
\item   For  off-plane regions, the lines of sight are not parallel to the
 $b=0^\circ$ and so the larger the distance the greater the height above the plane, and hence the lower the
 density. Therefore, the maximum density along the line of sight is 
closer than the real maximum in a plane parallel to $b=0^\circ$.
\end{itemize}

We can derive analytically the difference ($\Delta r=r_m-r_a$) between the
maximum density position along the line of sight ($r_m$) and the
intersection of the line of sight with the major axis ($r_a$) for a
triaxial structure with monotonically density decreasing  outwards.
The isodensity contours are defined by the points in space
with the same value of $t$:

\begin{equation}
t=\left(x_1^n+(x_2/A)^n+(x_3/B)^n \right)^{1/n},
\label{t}
\end{equation}
that in the n=2 case corresponds to isodensity ellipsoids. $A$ and $B$ are the axial ratios of the second and the third axes with
respect to the major axis of the ellipsoids,
and $x_i$ are the Cartesian coordinates with the axis of the ellipsoids
concentric with the centre of the Galaxy (see Fig. \ref{Fig:eliptan}). We assume that the minor axis is 
perpendicular to the Galactic plane ($x_3=z$). $x$, $y$, $z$ are the Cartesian coordinates with $XY$ defining
the plane of the Galaxy with the
$y$-axis in the Sun--Galactic Centre line and $\alpha $ is the angle between
the major axis of the ellipsoid and this $y$-axis (Fig. \ref{Fig:eliptan}).

The tangential point of the line of sight with the maximum density (filled square in Fig. \ref{Fig:eliptan}), follows:

\begin{equation}
\frac{\partial t}{\partial r}(r_m)=0
\label{der}
.\end{equation}

Equation \ref{der}, together with Eq.\ \ref{t}, lead to

\begin{equation}
x_1^{n-1}(r_m)\frac{\partial x_1}{\partial r}(r_m)+
\frac{x^{n-1}_2(r_m)}{A^n}\frac{\partial x_2}{\partial r}(r_m)+
\frac{x^{n-1}_3(r_m)}{B^n}\frac{\partial x_3}{\partial r}(r_m)=0
,\end{equation}

which finally lead to

\begin{equation}
\sum_{i=0}^{n-1}~a_i~r_m^i = 0,
\label{general}
\end{equation}
where

\[
a_{n-1}=(-1)^{n}\left(\cos^{n-1}(l+\alpha)+\frac{\tan(l+\alpha)
\sin^{n-1}(l+\alpha)}{A^n} \right)+
\]\begin{equation}
+ \frac{\tan^n b}{B^n~\cos(l+\alpha)},
\end{equation}
and

\[
a_i=(-1)^{i+1} \left(^{n-1}_{~~i~}\right)\left(\frac{R_\odot}{\cos
b}\right)^{n-1-i}\times
\]
\[ \times \left( \cos^{n-1-i}\alpha \cos^i(l+\alpha) + \frac{\sin^{n-1-i}\alpha \sin^i(l+\alpha)\tan(l+\alpha)}{A^n}\right),
\]
\begin{equation}
\forall i \in [0,n-2],
\end{equation}

while the position of the major axis (simple geometry with the application
of the sine rule, see Fig.\ \ref{Fig:eliptan}) is
\begin{equation}
r_a=\frac{R_\odot}{\cos b}\frac{\sin \alpha}{\sin (l+\alpha )}
.\end{equation}

\begin{figure}[!h]
\begin{center}
\vspace{1cm}
\mbox{\epsfig{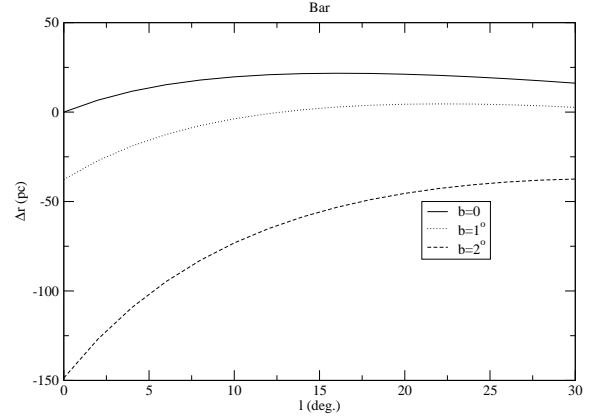}}
\end{center}
\caption{Difference between the
maximum density position along the line of sight for triaxial
ellipsoids of axial ratios 1:0.11:0.04, orientated with respect to the Sun--Galactic
Centre line at an angle of 43$^\circ $ (expected for the long bar) and the intersection of
their major axes with the line of sight.}
\label{Fig:calc_bar}
\end{figure}

In the simplest case, for $n=2$, Eq.\ \ref{general} yields

\begin{equation}
r_m=\frac{\frac{R_\odot}{\cos b}\left[\cos \alpha +
\frac{\tan (l+\alpha)\sin \alpha}{A^2}\right]}
{\cos (l+\alpha )+\frac{1}{A^2}\sin (l+\alpha )\tan (l+\alpha )+
\left(\frac{\tan b}{B}\right )^2\frac{1}{\cos (l+\alpha )}}.
\label{ecmax}
\end{equation}
Both of the expressions for $r_m$ and $r_a$ are coincident for $A\ll$ (very
elongated ellipsoids) and $\tan b\ll$ (in the plane), but  other cases 
 are affected by a significant systematic error $\Delta r$.
 Two triaxial structures were examined in order to estimate this effect:
\begin{itemize}
\item  $A=0.5$, $B=0.4$, $\alpha =28^\circ $, typical of the  bulge
(L\'opez-Corredoira et al. 2005): Fig. \ref{Fig:calc_bulge};
\item $A=0.11$, $B=0.04$, $\alpha =43^\circ $, typical of a long bar:
Fig.\ \ref{Fig:calc_bar}. The low depth in the line of sight of the bar (around 1 kpc) 
is justifiable because of the low dispersion of the red-clump
giants.
\end{itemize}
 The result is that $\Delta r$ is negligible for the bar,
with less than $\sim$$150$~pc of systematic error; however, it is not negligible
for the bulge, which reaches a discrepancy of up to 1000 pc.
 
\begin{figure}[!h]
\begin{center}
\vspace{1cm}
\mbox{\epsfig{file=6185fig30.ps,height=5.5cm}}
\end{center}
\caption{Difference between the
maximum density position along the line of sight for triaxial
ellipsoids of axial ratios 1:0.5:0.4, oriented with respect to the Sun--Galactic
Centre at an angle of 28$^\circ $ (expected for the thick bulge)
and the intersection of their major axes with the line of sight.}
\label{Fig:calc_bulge}
\end{figure}

To avoid this effect in deriving the real parameters of the bulge we performed a chi-squared fit
 to the observed distances of the red-clump density 
maxima by means of eq.\ (\ref{ecmax}), which contains only three free parameters ($A$, $B$ and $\alpha$). 
We have selected only 
those points at $|b|>3^\circ$, as they are more representative of this component. For each set of model 
parameters we calculate 
$\chi^2$. Once this was calculated for all sets in the parameter space, we select that one
 which gives a minimum $\chi^2$. Again, for $n=2$ the values which give the minimum $\chi^2$ are:

\begin{center}
$\alpha=13\fdg$3$^{+4.7}_{-2.3}$\\
$A=0.48^{+0.12}_{-0.13}$\\
$B=0.28^{+0.04}_{-0.03}$\\
\end{center}

The result for $\alpha$ is notably smaller than that directly  derived from the fit to the red-clump density peak,
 but it is consistent with other estimates found in previous works (Freudenreich 1998; L\'epine \& Leroy 2000; 
Picaud \& Robin 2004). The axial ratios 
of the bulge are more or less coincident with those
 considered in L\'opez-Corredoira et al.\ (2005) and others (e.g. Stanek et al. 1997; Bissantz \& Gerhard 2002; 
Picaud \& Robin 2004). 
It should be noted, however,  that eq.\ \ref{ecmax} is valid only for a series of triaxial  ellipsoids, 
and so will not be correct if the bulge has a different  geometry. However, when we use the general formula given by 
eq.\ \ref{general} the
results do not change noticeably (see Table \ref{tablachi}), all of them being 
compatible with a structure with a position angle of 
10--15$^\circ$ (with a weighted mean of 12$\fdg$6 $\pm$ 3$\fdg$2).

\begin{table}[!h]
 \caption[]{Parameters derived for different assumed bulge geometries}
\label{tablachi}
      \begin{center}
    \begin{tabular}{cccccc}
\hline
$n$ & $A$ & $B$ & $\alpha$ ($^\circ$) & $\chi_f^2$ & Type \\ 
\hline
\hline
2 & 0.48$^{+0.12}_{-0.13}$ & 0.28$^{+0.04}_{-0.03}$ & 13$\fdg3^{+4.7}_{-2.3}$  &
1.12 & ellipsoids \\
3 & 0.50$\pm$0.05 & 0.45$\pm$0.05 & 10$\fdg2^{+2.8}_{-1.5}$ & 1.36 & ---\\
4 & 0.55$\pm$0.03 & 0.45$\pm$0.03 & 14$\fdg2^{+3.0}_{-3.5}$  &  4.83 & boxy bulge\\
\hline
 \end{tabular}
\end{center}
\end{table}

In summary, the position angle we have derived in this work for the bulge (23$\fdg$1 $\pm$ 0$\fdg$9) by using  the
distribution of red-clump density maxima might be overestimating the real position angle by around 8--10$^\circ$. So 
the real position angle of the
inner bulge is around 13--15$^\circ$ with respect to the Sun--Galactic Centre line (nearly coincident with the value 
of 12$\fdg$6 $\pm$ 3$\fdg$2
obtained via chi-squared fitting). In the case of the long thin bar, the effect analysed
in this section is less significant and so do not alter the geometry of this component.

\section{Confirming the result using 2MASS star counts}
The 2MASS survey (Skrustkie et al.\ 2006) provides a complete coverage of the
Galactic Plane in the $JHK_{\rm s}$. Unfortunately, the  2MASS data
are not deep enough to reach the red-clump stars  in the
inner Galaxy, hence the 2MASS data  cannot be analysed using the above method. 
However, the  2MASS star counts can be
useful to confirm  the proposed  morphologies in the inner
Galaxy, as they are a powerful tool when searching for asymmetries in
the stellar distribution. While the bulge
is observed in off-plane regions up to $b\sim10^\circ$ and at
$|l|<15^\circ$, the long  bar is visible only in the
in-plane regions, $|b|<2^\circ$ and up to $l=27^\circ$ at positive
longitudes and but only to $l=-15^\circ$ at negative longitudes (from the length and position angle). 
For this study,  counts up
to $m_K$ = 9 mag will be used as this  gives a high  inner Galaxy-to-disc contrast (Garz\'on et al.\ 1993)
whilst still providing sufficient sources to give good statistics. A
similar analysis was performed in L\'opez-Corredoira et al.\ (2001) with
the combination of DENIS (Epchtein et al.\ 1997)
 and TMGS (Garz\'on et al.\ 1993) $K$-band counts. The advantage of using 2MASS is that it covers the whole region
 homogeneously and so makes testing the model simpler.

\begin{figure}[!h]
\centering
\includegraphics[width=6.75cm]{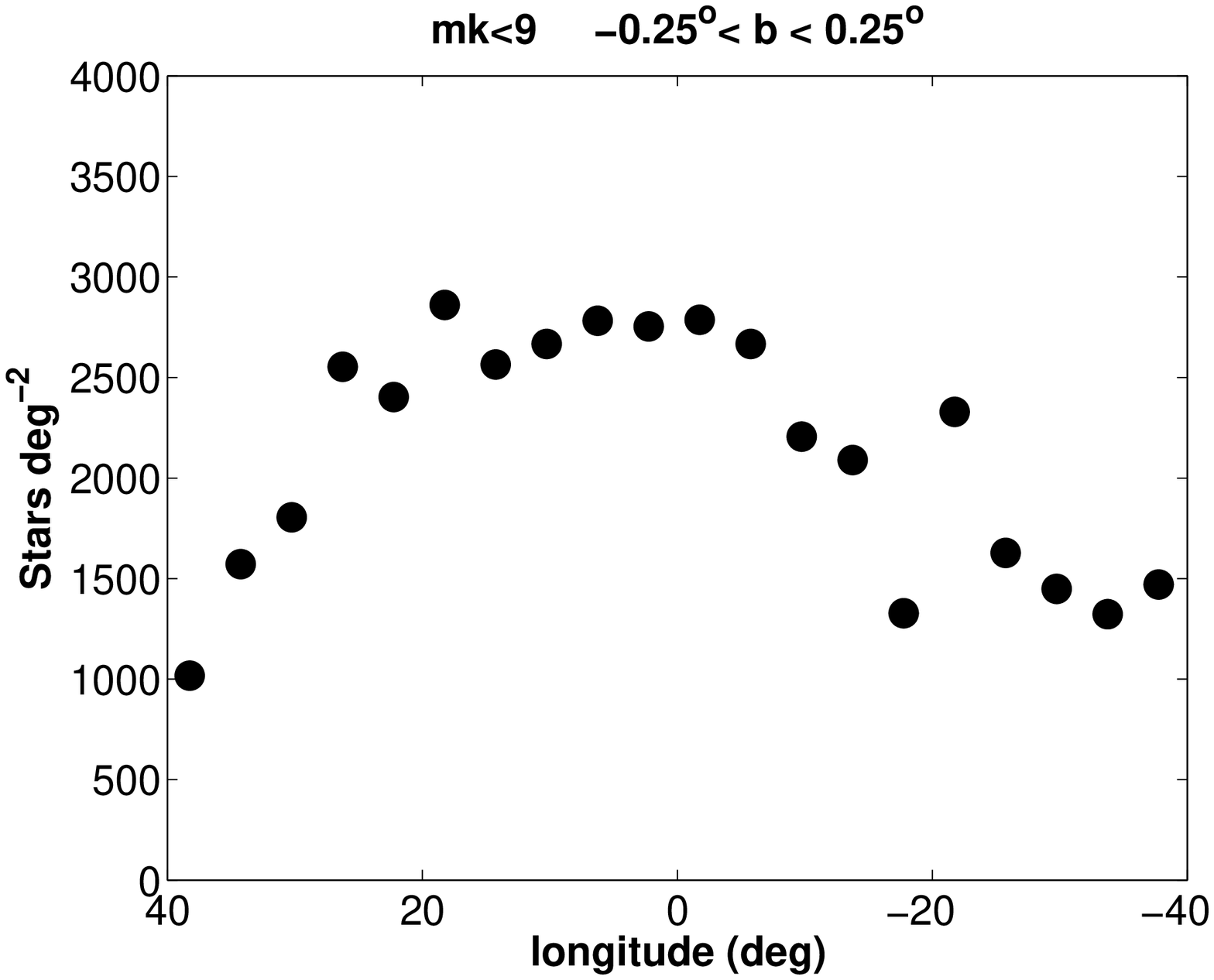}
\includegraphics[width=6.75cm]{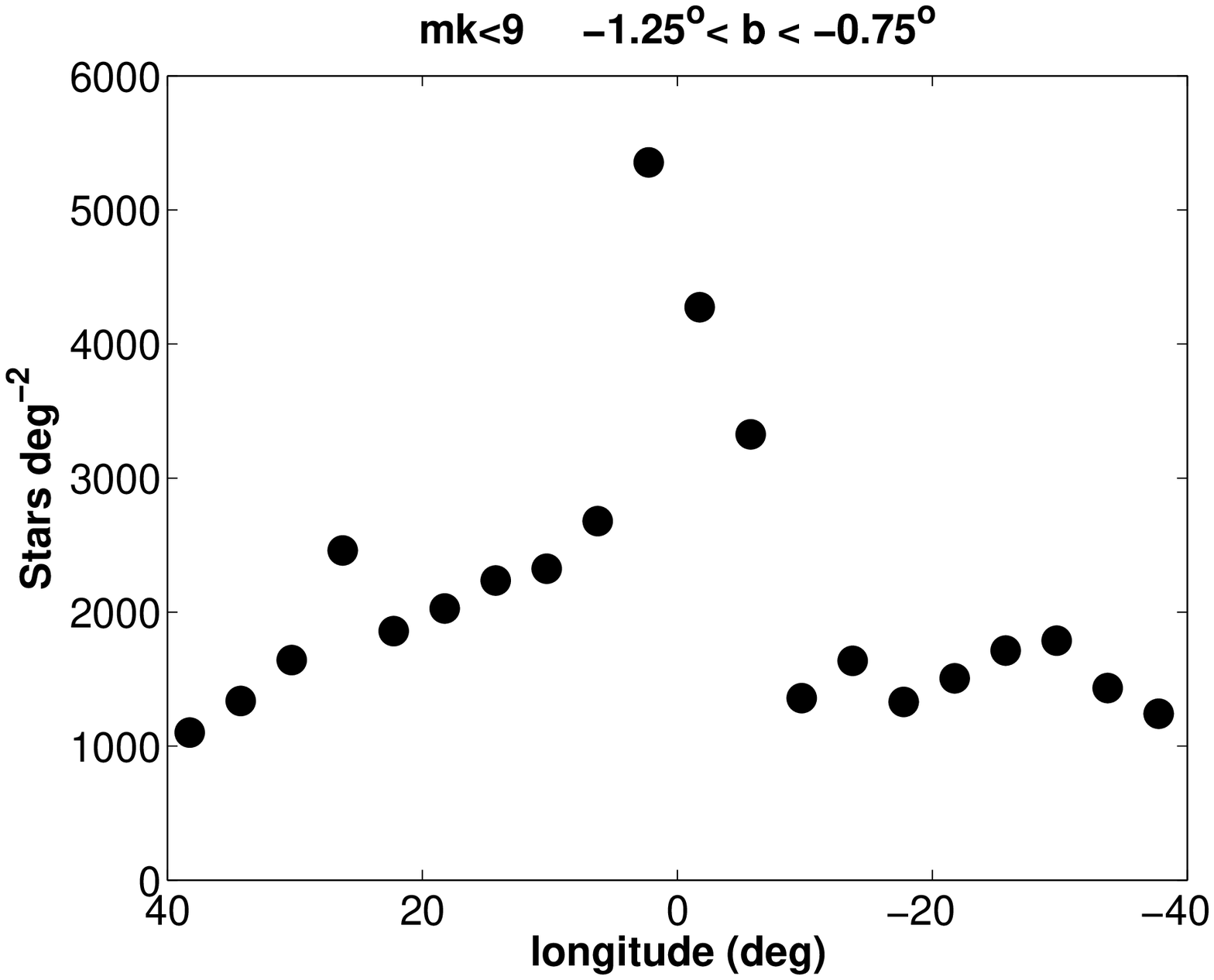}
\includegraphics[width=6.75cm]{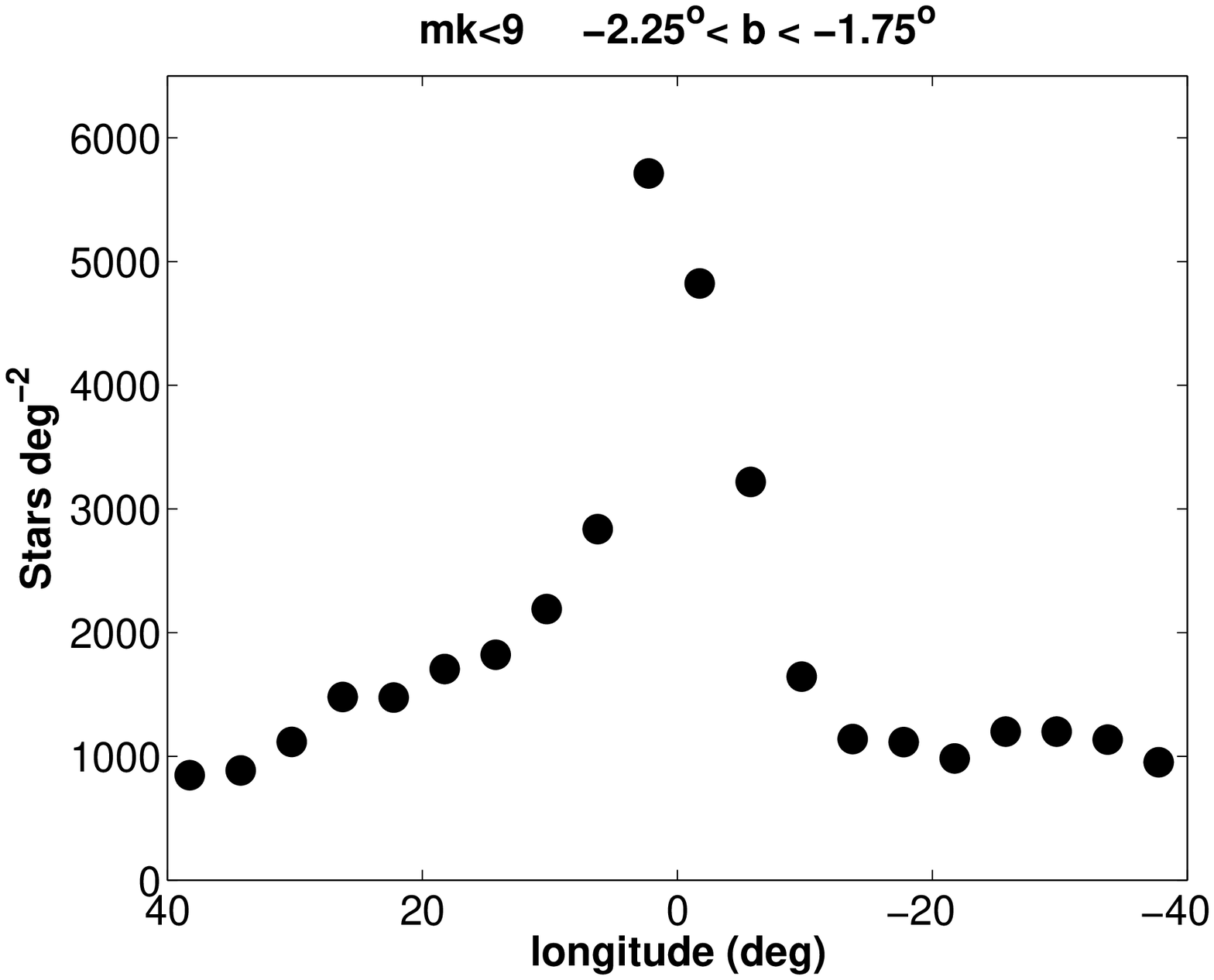}
   \caption{2MASS star counts with  $m_K < 9$ mag in three different latitude
ranges. Counts have been averaged over $\Delta l=5^\circ$ to obtain
   a large number stars in each region, in this way minimizing  the
Poissonian error.}
   \label{2masscounts}
\end{figure}

Figure \ref{2masscounts} shows the 2MASS star counts with $m_K < 9$ mag (averaged over $\Delta l=5^\circ$) at
three different latitude slices:
$-0.25^\circ < b < 0.25^\circ$,
$-1.25^\circ < b < -0.75^\circ$ and  
$-2.25^\circ < b < -1.75^\circ$. At the distance of the Galactic Centre (7.9~kpc), the heights above the Galactic plane 
 are 0 pc, $-$140 pc and $-$280 pc
respectively.

At $b=0^\circ$ there is a large-scale asymmetry about $l=0^\circ$  with far more counts at positive
longitudes than at negative longitudes. Farther from the plane  
young inner Galaxy features with very small scale heights should not be significant  and the effect of extinction 
is reduced. 
Hence, although the contrast with  the disc is reduced there should still be sufficient bar counts to see something and 
the accuracy of  model should be massively improved as it cannot deal with the patchy extinction on the plane. 
In L\'opez-Corredoira et al.\ (2001) it was shown that inner
Galactic
star counts are more or less symmetric in latitude, with small
differences attributed to the Sun being about 15 pc above the Galactic plane
(Hammersley et al.\ 1995) and differences in the extinction above and
below the plane. Here, we concentrate our analysis only
on those slices at negative latitudes and we   assume that the structure of the inner
Galaxy is more or less symmetric about the Galactic plane.

We have constructed a very simple Galactic model that includes the
contributions of the Galactic disc, a  bulge, and a long thin bar to
reproduce the 2MASS
counts at  $\langle b\rangle = -2.0^\circ$. For the Galactic disc we have used the
luminosity function of Eaton et al.\ (1984) and an exponential density for the outer Galaxy
with a scale height of 285~pc and a scale length of 2.1~kpc (L\'opez-Corredoira et al.\ 2002).
 The inner 4~kpc of the disc, however, has a  constant density (L\'opez-Corredoira et al.\ 2004). It has  
 has been shown that continuing the exponential disc to the Galactic Centre leads to a significant overestimation 
 of the   $K$-band  star counts (L\'opez-Corredoira et al.\ 2001; Benjamin et al. 2005). This result is consistent with 
the disc 
 being a Freeman type II disc, which are common in galaxies with long bars. 
  The triaxial bulge has ratios 1:0.5:0.4 and a position angle of the major axis with respect to the 
Sun--Galactic Centre line of 27$^\circ$ (L\'opez-Corredoira et al.\
2005). The bulge luminosity function and density law
have been also taken from this work. Finally, a simple bar model in
agreement with Hammersley et al.\ (2000) was added. The bar has a depth along the line of sight 
of 500 pc, a half-length of 4 kpc and a position angle of 43$^\circ$.
The distribution was assumed to be constant along the bar, but
exponential in height above the plane and the luminosity function used
was the same as for the disc although the density was then normalized
to make the total counts match those at $l=27^\circ$. In all the cases,
the extinction model used is as described in Wainscoat et al.\ (1992).

\begin{figure}[!h]
\centering
\includegraphics[width=8cm]{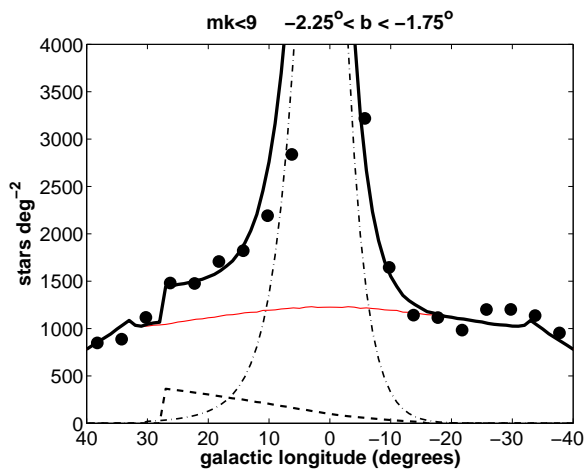}
\caption{2MASS star counts  to $K<9$ mag and the predictions of a model
that includes the contribution of disc (solid line), bulge (dot-dashed
line) and a long thin bar (dashed line). The sum of all the components
(strong solid line) reproduces the observed counts  very accurately.}
\label{counts}
\end{figure}

Whilst the model is simple, it can be shown from Fig.\ \ref{counts} that
it does reproduce the counts fairly well at $b =-2^\circ$. The model predicts: 
\begin{itemize}
\item The increasing counts with decreasing longitudes  for $|l| > 30^\circ$ due to the exponential disc
\item The sharp jump in the counts at $l=27^\circ$ due to the bar
\item The basically flat region  between $12^\circ < |l| < 27^\circ$ due to exponential disc  not continuing into the 
centre and the bar 
\item The steep rise inwards of $|l|<10^\circ$ as the bulge dominates. The bar/bulge ratio in 
this region is below 0.025 in this region
\end{itemize}
There are, however, some other features not
well predicted by the simple model. As no ring was included in the  model  there are excess counts 
for the region  $-30^\circ < l < -22^\circ$.
Sevenster et al.\ (1999) also  observed  an excess of OH/IR stars at $l=-22^\circ$. The possible implications 
for an elliptical ring are
 described in section 6.1 of L\'opez-Corredoira et al.\ (2001) but are beyond
the scope of this paper, so will not be repeated here. Similarly, the model does not include the Scutum 
spiral arm so near  the tangential point ($l=33^\circ$) there could also be problems. 
A fuller analysis of the inner Galactic star counts  can be found 
in L\'opez-Corredoira et al.\ (2001; section
3). Here, we are only interested in testing the plausibility of the
bulge + thin bar scenario when reproducing the observed counts.

It is, however, clear from the above that a disc + thin bar +  triaxial  bulge can
account for the a large part of the observed counts, in particular all of the large-scale Galactic features.
 Were the bar not to be included, the
fit would not be as good; in particular the asymmetry in longitude would not be reproduced.

\section{Conclusions}

 By analysing the distribution of red-clump sources along different lines of sight towards the inner Galaxy, we have 
demonstrated that the Milky Way has two different structures coexisting in the inner $R < 5$ kpc: a long thin bar with a
half length of 4 kpc constrained to the plane and a position angle of 43$^\circ$,
 and a thicker bulge with a position angle of 13--15$^\circ$.

In the TCS-CAIN data, each structure dominates  at different ranges of longitude and latitude. 
At $|b|>3.5^\circ$, $|l|<10^\circ$ the  bulge is the only component
that is observed whereas at  $20^\circ<l<28^\circ$ and $|b|\le0.5^\circ$ the long thin bar  dominates. The
scale heights of these structures are also very different. The red-clump giants of long-bar sources have a scale height 
 of $\sim$100 pc, 
whereas the disc is over 2.6 times  this  thick and the bulge sources
have a scale height $\sim$5 times larger still. Hence the long bar is geometrically a different structure to the 
triaxial bulge or disc,
although this does not necessarily mean that they are not related.

We have also checked how there are some systematic uncertainties in deriving distances from the red-clump 
population peaks that have not been taken into account in other
similar analyses (e.g.\ Stanek et al.\ 1994, 1997; Nishiyama et al. 2005; Babusiaux \& Gilmore 2005). 
Those uncertainties make the position angles
derived in those works around 10$^\circ$ larger than the real  ones. 

 We have shown that the large-scale  2MASS star counts in the inner Galaxy can be accurately described by a combination of a
  triaxial bulge, a bar and a disc. Hence, discussion as to the angle of the bar  should be  separated into two  
separate discussions:
  the form of the triaxial bulge and the form of the long bar.

\begin{acknowledgements}
   This publication makes use of 
   data products from TCS-CAIN (based on observations made at Carlos S\'anchez Telescope operated
by the IAC at Teide Observatory on the island of Tenerife) and 2MASS (which is
a joint project of the University of Massachusetts and the Infrared Processing
and Analysis Center (IPAC), funded by NASA and the NSF).

\end{acknowledgements}

\end{document}